\documentclass[12pt,reqno]{amsart}
\usepackage{amsmath,amssymb,graphicx,amscd,bm}
\usepackage{amsaddr}
\usepackage{mathrsfs}
\usepackage{enumerate}
\usepackage{mathrsfs}
\usepackage{dsfont}
\usepackage{amsfonts}
\usepackage{fullpage}

\usepackage[pagewise]{lineno} 

\usepackage{hyperref} 
\hypersetup{
    colorlinks=true,
    linkcolor=blue,
    filecolor=magenta,      
    urlcolor=magenta,
    }

\usepackage{caption}
\usepackage{subcaption} 

\usepackage{float} 
\usepackage{color} 

\usepackage{color}

\usepackage[utf8]{inputenc} 
\usepackage[T1]{fontenc}    
\usepackage{hyperref}       
\hypersetup{
    colorlinks=true,
    linkcolor=blue,
    filecolor=magenta,      
    urlcolor=cyan,
    pdftitle={Overleaf Example},
    pdfpagemode=FullScreen,
    }

\usepackage{url}            
\usepackage{hyperref}
\usepackage{booktabs}       
\usepackage{amsfonts}       
\usepackage{nicefrac}       
\usepackage{xcolor}         
\usepackage{float}          

\usepackage{amssymb}
\usepackage{placeins}
\usepackage{textcomp}
\usepackage{makecell}
\usepackage{mathrsfs}
\usepackage{wrapfig}
\usepackage{amsmath}
\usepackage{physics}
\usepackage{amsthm}
\usepackage{amsfonts}
\usepackage{comment}
\usepackage{mathtools}
\usepackage{amstext}
\usepackage{dsfont}
\usepackage{color}
\usepackage{booktabs}

\usepackage{enumerate}
\usepackage{paralist}
\usepackage{setspace}
\usepackage{scalerel}
\usepackage{graphicx}
\usepackage{caption}

\newcommand{\Rb}{\mathbb{R}}
\newcommand{\Nb}{\mathbb{N}}
\newcommand{\Pb}{\mathbb{P}}
\newcommand{\Qb}{\mathbb{Q}}
\newcommand{\Tc}{\mathcal{T}}

\newcommand{\1}{\mathds{1}} 

\newcommand{\Fc}{\mathcal{F}}
\newcommand{\Gc}{\mathcal{G}}
\newcommand{\Hc}{\mathcal{H}}
\newcommand{\Nc}{\mathcal{N}}

\newcommand{\EQ}[1]{\mathbb{E}^\mathbb{Q}\left[#1\right]}

\newtheorem{definition}{Definition}[section]

\usepackage{appendix}
\usepackage{algorithm}
\usepackage{algpseudocode}

\begin{document}

\title{Conditional Forecasting of Margin Calls using Dynamic Graph Neural Networks}

\author[1]{Matteo Citterio}
\address[1]{Dipartimento di Fisica, Universit\`{a} degli Studi di Milano,\\Milano, Italy.}
\email{matteo.citterio@studenti.unimi.it}

\author[2]{Marco D'Errico $\dagger$}
\thanks{$\dagger$The views expressed in this paper are those of the authors and do not necessarily reflect those of the ECB, the ESRB or its member institutions..}
\address[2]{European Systemic Risk Board Secretariat, European Central Bank,\\Frankfurt, Germany.}
\email{Marco.Derrico@ecb.europa.eu}

\author[3]{Gabriele Visentin*}
\thanks{*Corresponding author.}
\address[3]{Department of Mathematics, ETH Zurich, Zurich, Switzerland.}
\email{gabriele.visentin@math.ethz.ch}

\date{\today}



\begin{abstract}
We introduce a novel Dynamic Graph Neural Network (DGNN) architecture for solving conditional $m$-steps ahead forecasting problems in temporal financial networks. The proposed DGNN is validated on simulated data from a temporal financial network model capturing stylized features of Interest Rate Swaps (IRSs) transaction networks, where financial entities trade swap contracts dynamically and the network topology evolves conditionally on a reference rate. The proposed model is able to produce accurate conditional forecasts of net variation margins up to a $21$-day horizon by leveraging conditional information under pre-determined stress test scenarios. Our work shows that the network dynamics can be successfully incorporated into stress-testing practices, thus providing regulators and policymakers with a crucial tool for systemic risk monitoring.
 
\end{abstract}

\maketitle

\section{Introduction}

In the wake of the global financial crisis, one of the foundational objectives of regulatory reform was to ``make derivatives markets safer'' as articulated in the reforms proposed by the G20 leaders in 2009.\footnote{See, e.g., the \href{https://www.fsb.org/uploads/Overview-of-Progress-in-the-Implementation-of-the-G20-Recommendations-for-Strengthening-Financial-Stability.pdf}{Financial Stability Report report to G20 Leaders (2014)} and \href{https://www.fsb.org/uploads/P030717-2.pdf}{Implementation and Effects of the G20 Financial Regulatory Reforms (2014).}} These reforms targeted the reduction of systemic risks inherent in derivatives by promoting central clearing, margin requirements, and enhanced transparency. 

Despite these efforts, derivatives markets have faced recurrent instability in recent years, suggesting that vulnerabilities remain. Events such as the Nasdaq Clearing member failure, the COVID-19 ``dash for cash'' in March 2020, nickel market volatility in March 2022, the European energy crisis, and the UK gilt market turmoil in October 2022 underscore the continued susceptibility of these markets to systemic risk. These episodes have exerted substantial pressure on market infrastructures, at times causing spillovers into the broader economy and, in certain cases, necessitating public interventions to stabilize the financial system. The scale of margin calls and participants' ability to meet them have become central concerns.\footnote{See the \href{https://www.bis.org/bcbs/publ/d537.htm}{BCBS-CPMIIOSCO Review of margining practices (2022)}.} While margin requirements mitigate counterparty credit risk, they also introduce challenges like liquidity pressures and increased procyclicality -- issues evident in recent stress episodes.

Significant advancements in data collection and analysis for systemic risk monitoring have enabled near real-time mapping of the full network of margin calls between counterparties in derivatives markets.\footnote{See 
\href{https://www.esrb.europa.eu/pub/pdf/reports/esrb.report200608_on_Liquidity_risks_arising_from_margin_calls_3~08542993cf.en.pdf}{Liquidity risks arising from margin calls, ESRB (2020)}.} However, despite these efforts, current approaches remain largely \textit{ex post}, underscoring the need for a proactive, system-level forecasting network framework. This raises a key research question: to what extent can a comprehensive \textit{ex ante} framework be developed to quantify margin calls in stressed derivatives markets, leveraging the extensive transaction-level data available to authorities?

There is a rich literature focusing on incorporating network structure in systemic risk management. Starting with the pioneering work by Eisenberg and Noe \cite{eisenberg2007systemic}, numerous authors have studied the impact of shock propagation on financial networks and investigated the dependence of systemic risk on network connectivity \cite{upper2011simulation, battiston2012default, glasserman2015likely, acemoglu2015systemic} and the problems related to its accurate estimation \cite{battiston2016price}.

Most works have traditionally focused on the existence of clearing interbank payments under increasingly realistic contagion models, including equity cross-holding and arbitrary seniority structures \cite{elsinger2009financial}, hard defaults  \cite{gai2010contagion}, bankruptcy costs \cite{rogers2013failure}, liquidity cascades \cite{gai2010liquidity, lee2013systemic}, asset fire sales \cite{cifuentes2005liquidity, caccioli2014stability}, funding liquidity cascades \cite{diamond2000bank, amini2012mathematical}, feedback effects in bilateral credit valuation adjustments \cite{battiston2016debtrank}, and multiple maturities \cite{kusnetsov2019interbank}. Some authors have also explored the issues surrounding derivative transactional networks by studying the optimal design of central clearing counterparties \cite{amini2015systemic}, compression of OTC markets \cite{d2021compressing}, and bilateral credit valuation adjustment for CDS portfolios \cite{bo2014bilateral}. All the models cited above assume the network of contracts to be static during the contagion propagation, thus excluding from the analysis important sources of contagion related to the dynamic behavior of financial entities. A few recent works have attempted to integrate a dynamic component by introducing stochastic external assets in both the discrete-time \cite{feinstein2022endogenous, feinstein2021dynamic, calafiore2023clearing, calafiore2022control} and the continuous-time settings \cite{banerjee2024dynamic, sonin2020continuous, chen2021financial, barucca2020network}, but crucially they retain the assumption of a deterministic liability matrix. 

Our work, instead, is a contribution to systemic risk management on fully dynamic financial networks, by which we mean temporal networks where financial entities can dynamically exchange bilateral contracts in an evolving stochastic environment. Other authors have recently explored this setting, by addressing the problem of link prediction in a temporal interbank transaction network using a mixed GCN-LSTM architecture \cite{zhang2022deep}, using a latent space dynamic model for monthly interbank networks \cite{linardi2020dynamic} and modelling temporal transactional data using interacting measure-valued processes \cite{capponi2020dynamic}.

Given the increasing availability of regulatory OTC transactional data, as discussed above, and inspired by the work of Zhang \cite{zhang2022deep}, we propose a new Dynamic Graph Neural Network (DGNN) architecture specifically designed to solve conditional $m$-steps ahead forecasting problems in temporal financial networks. The goal of our model is to provide accurate conditional forecasts of systemically relevant node features by leveraging conditional information under pre-determined macro-economic stress test scenarios. We train and validate the proposed DGNN on simulated data from a temporal financial network model capturing stylized features of Overnight Indexed Swaps (OIS) transaction networks, where financial entities trade OIS contracts dynamically and the network topology evolves conditionally on the OIS reference rate. Our work shows that the network dynamics can be successfully incorporated into stress test practices, thus providing regulators and policymakers with a crucial tool for systemic risk monitoring.

The rest of the paper is organized as follows: Section \ref{sec:background} presents some background on DGNNs, Section \ref{sec:problem} introduces our dynamic simulation model and formalizes the conditional forecasting problem we want to solve, Section \ref{sec:model} introduces our proposed DGNN architecture and Section \ref{sec:experiments} discusses its performance, finally Section \ref{sec:conclusion} concludes.

\section{Background on Dynamic Graph Neural Networks (DGNNs)}
\label{sec:background}

In the machine learning community there is increasing interest in structured problems on temporal networks \cite{kazemi2020representation, skarding2021foundations, feng2024comprehensive}. From a machine learning perspective, this kind of data retains the well-known exploitable features of static graphs \cite{hamilton2020graph} while incorporating a temporal evolution \cite{Wu}. 

Dynamic Graph Neural Networks (DGNNs) are extensions of Graph Neural Networks (GNNs) to temporal graphs, which leverage latent dynamic representations of the data to perform prediction tasks \cite{dynamicGNNs_review_1}.

Within the domain of DGNNs, a distinction needs to be made between continuous-time networks and discrete-time networks, operating respectively on event-based representations and snapshot-based representations of temporal networks. In this work, given the discrete-time nature of the forecasted variables, we focus on the snapshot-based representation, with a fixed node set over the time horizon.

The basic approach in discrete-time DGNNs is to encode the network snapshot by snapshot, updating a latent representation of the dynamic graph at each step. This is achieved by employing the neighborhood aggregation technique at each snapshot, as normally done by static GNNs, such as Graph Convolution Networks (GCNs), combined with a dynamic evolution step typically handled by a Recurrent Neural Network (RNN) or a Long-Short Term Memory (LSTM) network \cite{gcrn}. 

\begin{figure}[h]
    \centering
    \includegraphics[width=0.9\textwidth]{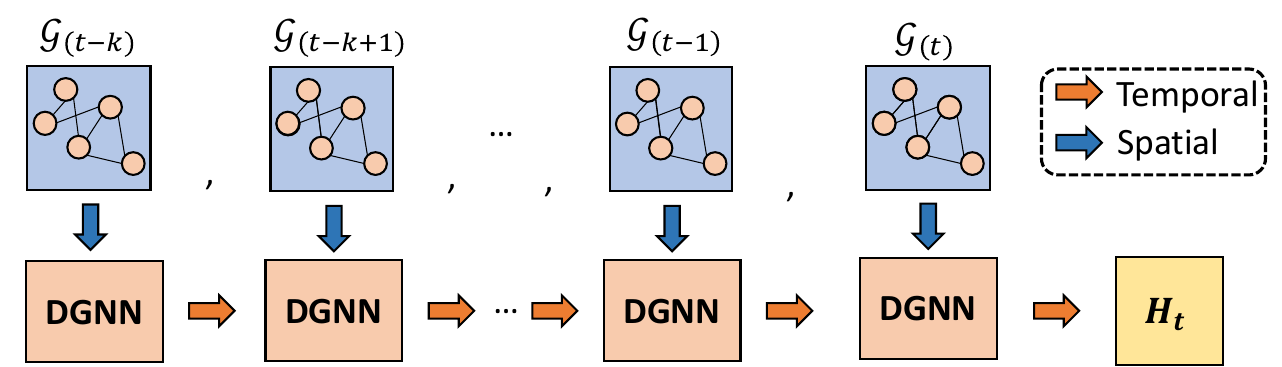}
    \caption{Learning framework of a typical discrete-time DGNN.}
    \label{fig:general_dgnn_framework}
\end{figure}

\begin{figure}[h]
    \centering
    \includegraphics[width=0.9\textwidth]{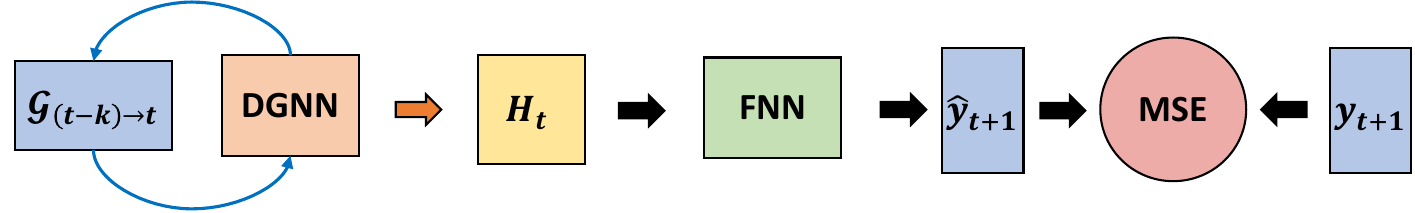}
    \caption{Encoder-decoder framework for DGNNs.}
    \label{fig:unrolled_dgnn}
\end{figure}

\paragraph{Encoder-decoder framework and model training.} DGNNs adhere to the encoder-decoder framework of classical GNNs \cite[Section 3.1]{hamilton2020graph}. Under this setting, a DGNN can be seen as an encoder which processes dynamical graph data and outputs a node \emph{embedding} or \emph{latent representation} $H_t \in\Rb^{N\times C}$ (where $C$ denotes the latent space dimension) at each time step $t$ based on the previous $k$ time-steps. In creating this embedding, the model is required to learn both spatial and temporal patterns in the sequence (see Figure \ref{fig:general_dgnn_framework}. Subsequently, this learned representation can be used by a decoder - typically a simple Feed-Forward Neural Network (FFNN) with one or two hidden layers - to perform various prediction tasks \cite{dygrencoder}, as illustrated in Figure \ref{fig:unrolled_dgnn}. Depending on the specific prediction task, the final nonlinearity may be a \textit{softmax} function (e.g., for classification problems) or a simple \textit{sigmoid} (e.g., for node regression and link prediction). 

\begin{figure}[h]
    \centering
    \includegraphics[width=0.9\textwidth]{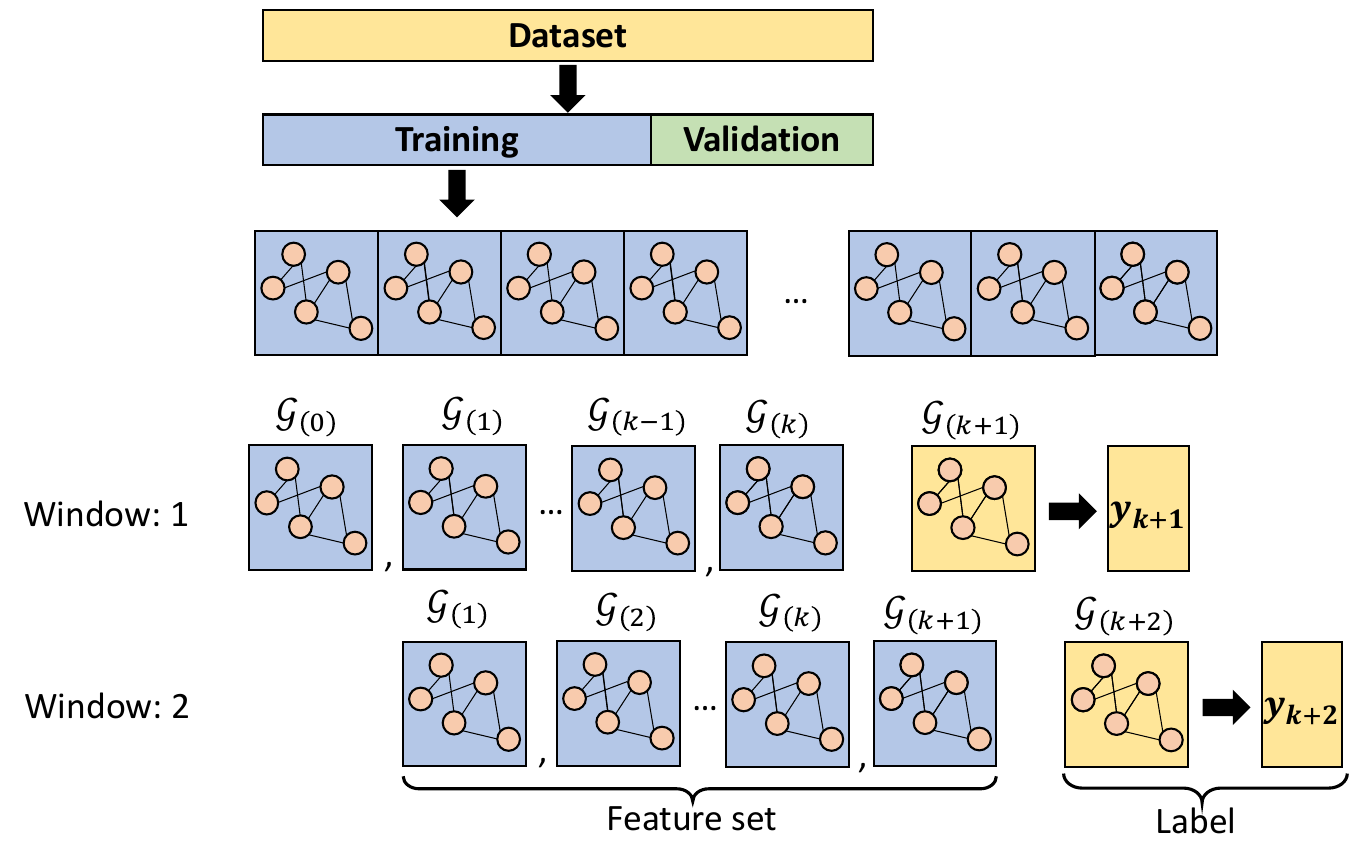}
    \caption{Data windowing process for a discrete-time DGNN.}
    \label{fig:data_windowing}
\end{figure}

The training methods for discrete DGNNs are those typical of recurrent architectures \cite{ConvLSTM}. In other words, it is common to use \textit{data windowing} to segment the sequential input data. This process, depicted in Figure\ref{fig:data_windowing}, generally involves two steps. Initially, the dataset is divided into training and validation sets. Subsequently, each of them is subdivided into overlapping windows of a predefined length $k$. Finally, each window is associated with a label, which is task-dependent and obtained from the $m$-th subsequent snapshot of the sequence. For example, in a typical single-step ($m=1$) node regression problem, as depticted in Figure \ref{fig:data_windowing}, the label of the first window consists of a node feature vector $y_{k+1}\in\Rb^{N}$ extracted from the snapshot at time $t=k+1$ \cite{dynamicGNNs_review_1}.

In the following sections, we introduce two representative models embodying a slightly different strategies for combining the GNN component with the RNN one. Specifically, the two approaches differ in how temporal dependencies are captured. The first method, exemplified by the GC-LSTM model, involves using the latent representation generated by the GNN as an input for an RNN, whereas the EvolveGCN uses an RNN to learn the temporal evolution of the GNN's weights.

\paragraph{GC-LSTM.} The most straightforward approach is to use a GNN to process each snapshot of the graph and feed its output to a time series component, such as an RNN \cite{dynamicGNNs_review_1}. This is precisely the case of GC-LSTM \cite{gc-LSTM}, which combines multiple GCN layers \cite{kipf_gcn} with an LSTM \cite{LSTM}. However, several other works have also adopted this approach \cite{ConvLSTM, gcrn}.

GC-LSTM builds upon the success of LSTMs by performing convolution operations on the cell layer state and the hidden layer state at the previous time step. This is accomplished using four different graph convolution networks, each dedicated to one of the gates of the LSTM layer. Overall, the model is described by the following set of equations:

\begin{equation}
\begin{split}
    i_t &= \sigma\left( X_tW_{i} + \textrm{GCN}_i^k(\skew{5}\tilde{A}_{t-1},h_{t-1}) + b_i \right)\\
    f_t &= \sigma\left( X_tW_{f} + \textrm{GCN}_f^k(\skew{5}\tilde{A}_{t-1},h_{t-1}) + b_f \right)\\
    c_t &= f_t\odot c_{t-1} + i_t \odot \tanh(X_tW_{c} + \textrm{GCN}_c^k(\skew{5}\tilde{A}_{t-1},h_{t-1})+b_c)\\
    o_t &= \sigma\left( X_tW_{o} + \textrm{GCN}_o^k(\skew{5}\tilde{A}_{t-1},h_{t-1}) + b_o \right)\\
    h_t &= o_t \odot \tanh(c_t)
\end{split}
\end{equation}

Here, $\textrm{GCN}^k(A,X)$ denotes a $k$-layer Graph Convolutional Network operating over the (normalized) adjacency matrix $A_t$ at time t, while $X_t$ represents the node feature matrix. Although specifically conceived for dynamic link prediction, this model can serve as a general encoder for discrete temporal networks across a wide range of prediction tasks.

\paragraph{EvolveGCN} EvolveGCN \cite{evolvegcn} directly integrates a recurrent model into a GCN by evolving the weights of the GCN network through an LSTM or, occasionally, a GRU \cite{GRU}. At its core, the Evolving Graph Convolution unit updates the model's hidden states treating the weights $W$ as the hidden layer of the RNN, as follows:
\begin{align}
    W_t^{(l)} &= \textrm{GRU}\left(H_t^{(l)}, W_{t-1}^{(l)}\right)\\
    H_t^{(l+1)} &= \textrm{GNN}\left(A_t, H_{t}^{(l)}, W_t^{(l)}\right)
\end{align}
Here, the superscript $l$ denotes the layer of the network, where we set $H_t^{(l=0)} = X_t$. EvolveGCN is expected to train rapidly on an evolving network with minimal changes between snapshots \cite{dynamicGNNs_review_1}.

\begin{figure}[h]
    \centering
    \includegraphics[width=0.9\textwidth]{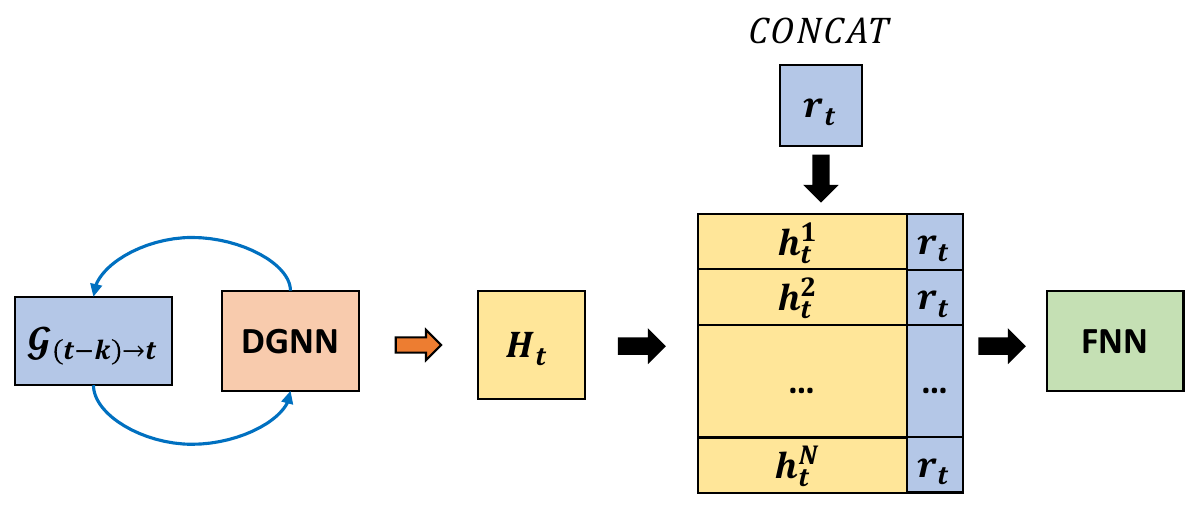}
    \caption{Conditional forecasting within the encoder-decoder DGNN framework.}
    \label{fig:conditioned_prediction}
\end{figure}

\paragraph{Conditional forecasts.} Within the encoder-decoder framework, one may be interested in performing \textit{conditional} forecasts, where a conditioning variable provides relevant information for the task at hand. For instance, DGNNs have been employed in time series generation conditioned on an auxiliary variable representing patient-specific information \cite{rcgan}. Our work is specifically tailored to solve such conditional forecasting problems and we follow the common approach of concatenating the conditioning variable to the latent representations generated by the DGNN before passing them to the decoder FFNN, as illustrated in Figure \ref{fig:conditioned_prediction}.

\paragraph{Multi-step ahead prediction.} An important application of a DGNN model is to generate good representations for forecasting a given quantity (at the node, edge or graph level) several time steps ahead in the future. The Sequence-to-Sequence \cite{Seq2Seq} modelling framework has emerged as a promising approach to handle this problem requiring only minimal variations on the single-step ahead forecasting problem.

The idea involves using an additional RNN or LSTM to produce a prediction of the latent representation at each subsequent future time step up to the final forecasting time and then using it as an input to predict the quantity of interest. This simple extension of the single-step ahead prediction setting has yielded remarkable results, mainly due to the fact that all models involved are trained simultaneously back-to-back, thus facilitating learning exactly those graph representations that have the greatest predictive power at the required future time step.

\begin{figure}[h]
    \centering
    \includegraphics[width=0.9\textwidth]{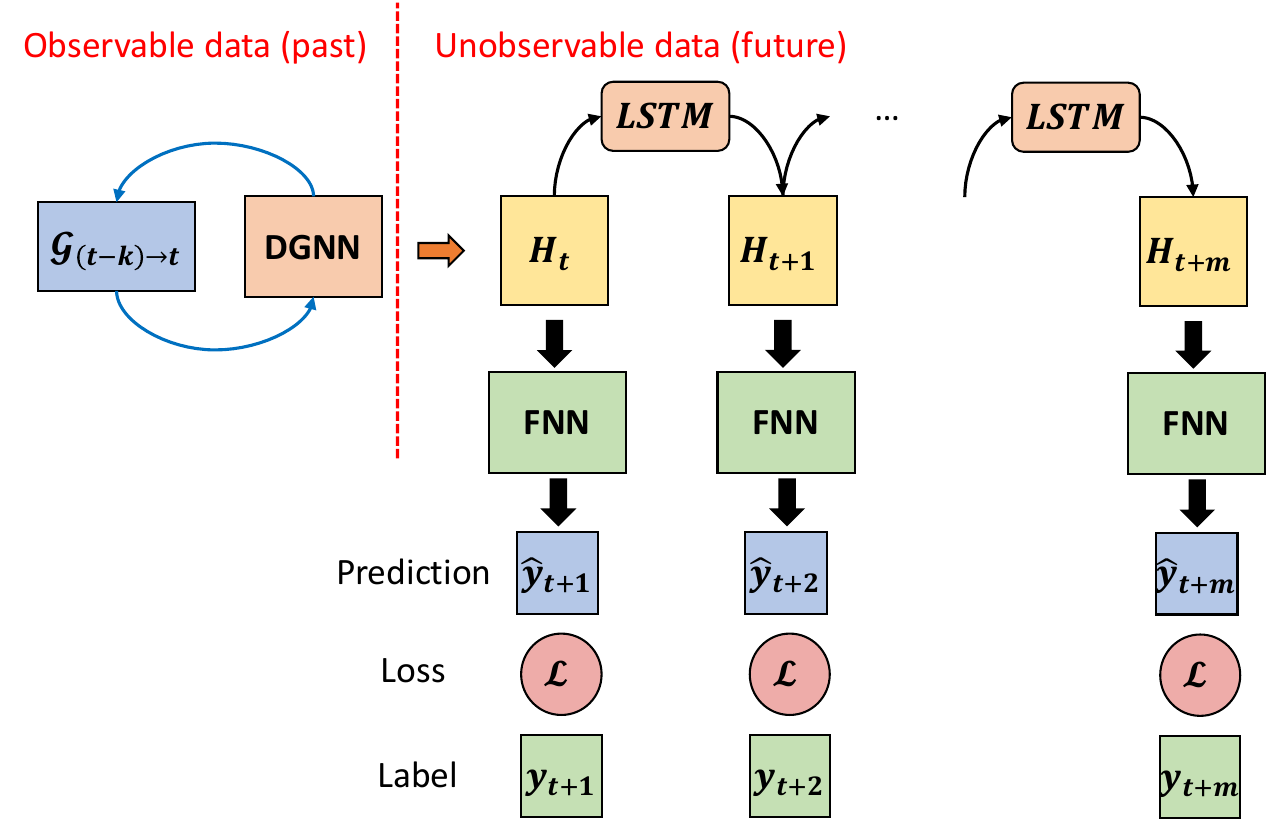}
    \caption{Seq2Seq setup for $m$-steps ahead prediction.}
    \label{fig:seq2seq}
\end{figure}

In practice, for a typical $m$-steps ahead prediction the same FFNN is used to produce a prediction at each intermediate step and the intermediate prediction losses are aggregated (typically using the mean) so that parameters can be updated through back-propagation and the optimizer of choice, as illustrated in Figure \ref{fig:seq2seq}. This procedure tends to result in models with better generalization.

The main drawback of this framework is certainly the computational cost, as the model becomes larger with each added future time step.

\section{Problem description}
\label{sec:problem}

In this section we introduce the precise setting of our multi-step ahead forecasting problem. First we present a synthetic dynamic network model in which nodes are financial entities and edges are dynamically traded Overnight Indexed Swaps (OISs), which are interest rate swaps indexed to a reference overnight rate, such as SOFR in the US, ESTER in the EU and SONIA in Switzerland \cite[Section 7.1]{hull_options}.

\begin{figure}[h]
    \centering
    \includegraphics[width=0.9\textwidth]{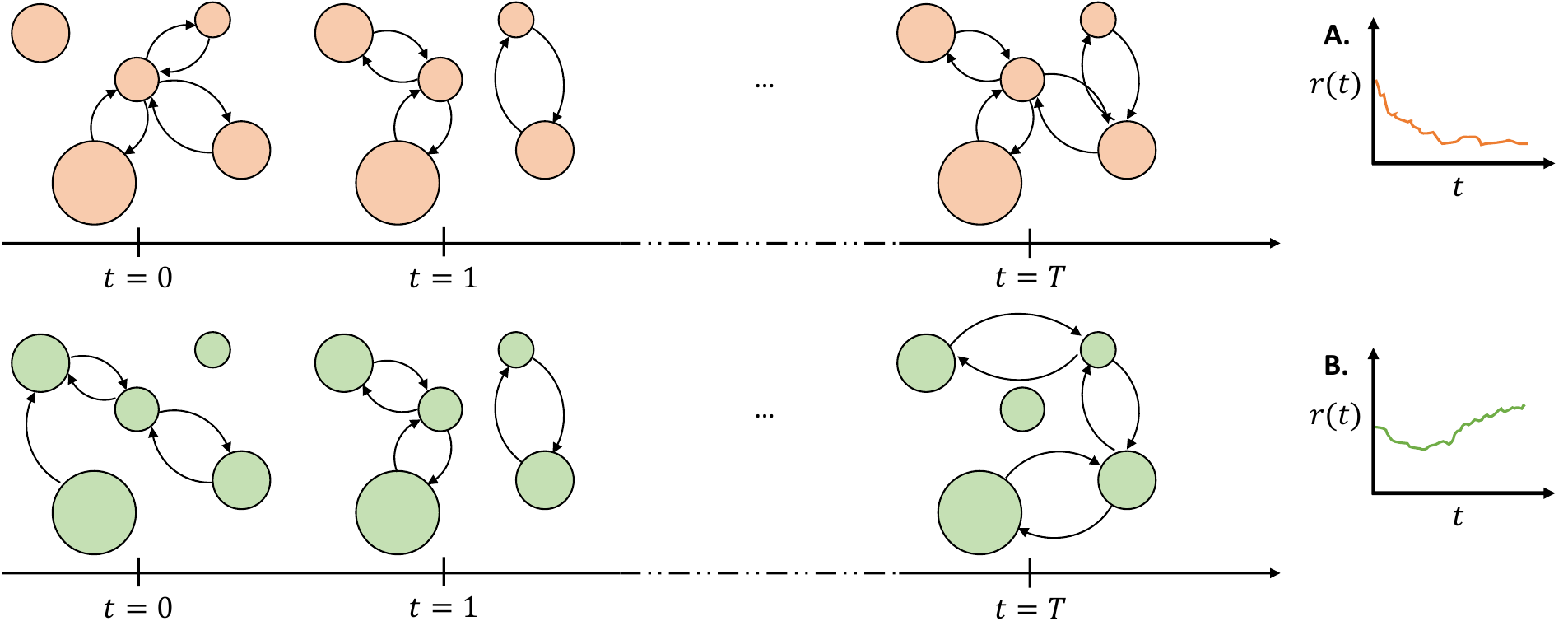}
    \caption{The network dynamics is influenced by the OIS reference rate.}
    \label{fig:pattern_of_shock}
\end{figure}

The trading behavior of the financial entities - and therefore the network dynamics - depends on the OIS reference rate, as shown in Figure \ref{fig:pattern_of_shock}, and the financial status of the entities themselves. 

In the following paragraphs, we detail how we model the OIS reference rate, the pricing of OIS contracts and the network dynamics.

\paragraph{Overnight reference rate.} Since overnight rates have been adopted as reference rates of OTC contracts only recently, there is not enough data availability to train a model on historical data. We have therefore chosen to simulate the reference rate from a stochastic model.

We work on a filtered probability space $(\Omega, \mathcal{A}, (\mathcal{A}_t)_{t \ge 0}, \mathbb{Q})$ satisfying the usual conditions \cite[Definition 2.25]{karatzas2014brownian} and model the reference rate $r(t)$ directly under a martingale measure $\Qb$. 

As common in interest rate theory \cite{filipovic2009term, brigo2001interest}, we model the reference rate $r(t)$ using an affine short-rate diffusion model, specifically a CIR process, with following dynamics:
\begin{equation}
    \begin{cases} d r(t) = \kappa (\theta - r(t)) dt + \sigma \sqrt{r(t)} d W_t,\\ r(0) \in \Rb_+ \end{cases} 
\end{equation}

where $\{W_t\}_{t \ge 0}$ is a Brownian motion under $\Qb$ and we assume that $2 k \theta \ge \sigma^2$ (to guarantee that $r(s) > 0$ for all $s > 0$ $\Qb$-a.s.). For a comprehensive description of the CIR process and corresponding simulation techniques, see \ref{subsec:interest_rate_process}. 

Since overnight rates are published daily and variation margins on OTC derivatives are computed daily, we naturally work in a discrete-time framework and fix a uniform time grid $\Tc = \{t_i, i\in\Nb\}$ of daily observations following the actual/365 convention without holidays, i.e. $\Delta t := t_i - t_{i-1} = 1/365$, for all $i$.

The rate $r(t_i)$ then represents the overnight rate on the $i$-th day. For instance, if the reference rate is SOFR, then $r(t_i)$ is the rate as announced at 8.00 AM ET on day $i$ from the aggregated transactions conducted in the overnight period $(t_{i-1},t_i]$.

Since overnight interest rates are compounded daily, we define the zero-coupon bond price at time $t$ with maturity $T$ as follows:

\begin{equation}
    p(t, T) = \EQ{ \prod_{t_i \in (t, T] \cap \Tc} \left( 1 + r(t_i)(t_i - t_{i-1})\right)^{-1} \bigg| \Fc_t},
\label{eq:bond_price_def}
\end{equation}

and we denote by $\Tc \ni T \mapsto R(t, T)$ for $t \in \Tc$ the corresponding daily term structure of simple spot rates, defined as:

\begin{equation}
    R(t, T) = \frac{1}{(T - t)} \left( \frac{1}{p(t, T)} - 1 \right)
\end{equation}

\paragraph{Overnight Indexed Swaps (OIS).} An Interest Rate Swap (IRS) is a swap contract where interest at a predetermined fixed rate, applied to a certain principal, is exchanged for interest at a floating reference rate, applied to the same principal, with regular exchanges being made for an agreed period of time \cite{hull_options}. If the reference rate is an overnight rate, then the swap contract is called an Overnight Indexed Swap (OIS). OIS contracts have become much more common after the phasing out of the LIBOR reference rate.

Denote by $N$ the principal of the OIS contract, by $t_0$ its start date, by $T$ its maturity, and by $K$ the predetermined fixed rate, then the value at time $\Tc \ni t \geq t_0$ of the fixed leg of the contract is the discounted expected value of all future fixed rate payments:

\begin{align}
    V^{\text{fixed}}(t) & = \EQ{ N \prod_{t_i \in (t, T] \cap \Tc} \left( 1 + r(t_i)dt_i\right)^{-1} \left( 1 + K(T - t_0) \right) \bigg| \Fc_t} \nonumber \\
& = N p(t, T) \left( 1 + K(T - t_0) \right). \label{eq:fixed_leg_t}
\end{align}

Analogously, the value of the floating leg at times $\Tc \ni t\geq t_0$ is the expected discounted value of all future floating rate payments:

\begin{align}
    V^{\text{float}}_t & = \EQ{ N \prod_{t_i \in (t, T] \cap \Tc} \left( 1 + r(t_i) \Delta t \right)^{-1} \prod_{t_i \in \Tc \cap (t_0, T]} \left( 1 + r(t_i) \Delta t \right) \bigg| \Fc_t} \nonumber \\
    & =N\cdot \EQ{\prod_{t_i \in (t_0, t] \cap \Tc}  \left( 1 + r(t_i) \Delta t \right) \bigg| \Fc_t} \nonumber \\
    & = N \prod_{t_i \in (t_0, t] \cap \Tc} \left( 1 + r(t_i) \Delta t \right) \label{eq:floating_leg_t}
\end{align}

The arbitrage-free value of the constant rate $K$ (also known as the \textit{fair} OIS rate) is then obtained by enforcing equality of Eq. \eqref{eq:fixed_leg_t} and Eq. \eqref{eq:floating_leg_t} at time $t_0$, i.e. at the start of the contract, resulting in: 
\begin{equation}
    K = \frac{1}{T - t_0} \left( \frac{1}{p(t_0, T)} - 1 \right) = R(t_0, T).
\label{eq:OIS_rate_definition}
\end{equation}

This value of $K$ is precisely the value of the fixed interest rate for all OIS contracts starting at time $t_0$.

The value of the OIC contract at time $t$, denoted by $V(t)$, is simply the difference between the values of two legs:
\begin{align}
    V(t) &= V^{\text{fixed}}(t) - V^{\text{floating}}(t) \nonumber \\
    & = \delta N \left( p(t, T) \left( 1 + K(T - t_0) \right) - \prod_{t_i \in (t_0, t] \cap \Tc} \left( 1 + r(t_i)(t_i - t_{i-1})\right)\right) \label{eq:OIS_value} \\ 
& = \delta N \left( \frac{1 + R(t_0, T)(T - t_0)}{1 + R(t, T)(T - t)} - \prod_{t_i \in (t_0, t] \cap \Tc} \left( 1 + r(t_i)(t_i - t_{i-1})\right) \right), \nonumber 
\end{align}

where $\delta$ denotes the receiver-payer indicator function, i.e. $\delta=+1$ for the fixed rate \textit{receiver}, $\delta=-1$ for the fixed rate \textit{payer}.

\begin{figure}[h]
    \centering
    \includegraphics[width=0.9\textwidth]{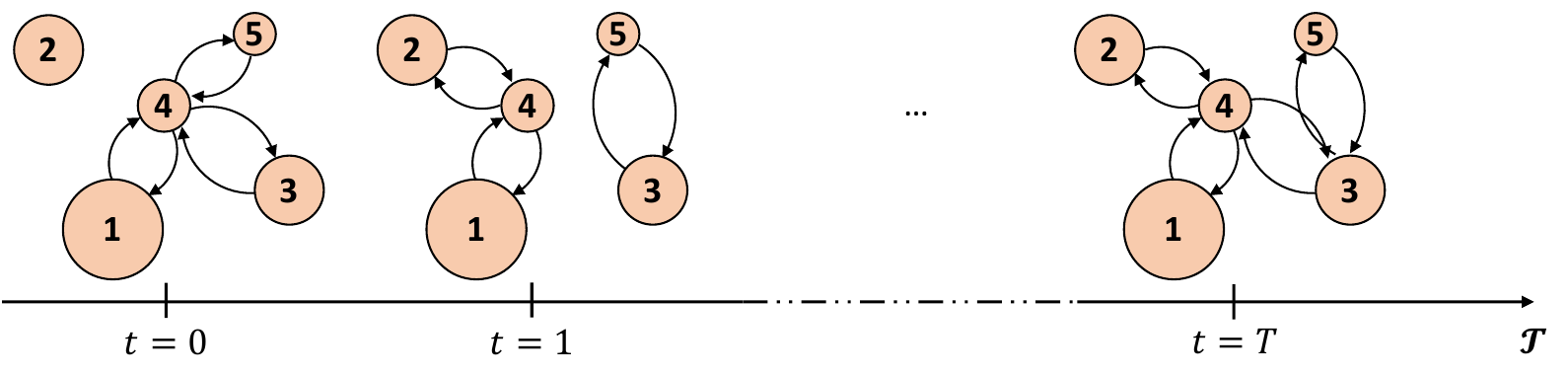}
    \caption{One realization of the proposed dynamic financial network model.}
    \label{fig:temporal_graph_example}
\end{figure}

\paragraph{Dynamic financial network model.} The financial network model we propose is a stochastic temporal network 
\begin{equation}
    \Gc=\left\{G_t=(V, X_t, E_t)\right\}_{t\in\Tc} 
\end{equation}
defined on a grid of daily observations, as shown in Figure \ref{fig:temporal_graph_example}.

The node set $V$ consists of $N$ financial entities and is partitioned into two classes: \emph{hubs}, representing big players in the OTC market, such as market makers and liquidity providers, and \emph{privates}, representing smaller entities, typically trading to hedge real risks or similar business reasons.

The matrix $X_t \in \Rb^{N\times C}$ contains the node features of the network at time $t$, where $C$ is the node feature dimension. Node features can represent information specific to each financial entity, such as its balance sheet or legal information. These features can affect the likelihood of entering a contract or the type of contract entered (e.g. as a fixed rate payer or as a fixed rate receiver). For the purposes of this synthetic model we limit ourselves to fixed binary features, therefore $X$ is simply a vector with components $X_i = +1$ for hubs and $X_i=-1$ for privates, but we emphasize that the proposed DGNN model is designed to be trained also on temporally evolving node features.

Since both the node set and the feature matrix $X$ are constant in time, the entire dynamics of the system is captured by the edge set $E_t$, which consists of OIS contracts between the financial entities. Edges can therefore appear stochastically in time and disappear at the contract's maturity. Since any pair of entities may have multiple outstanding contracts at a given time, we are dealing here with a dynamic \emph{multigraph} model.

We model the appearance of edges using a time point process. More specifically, for each pair of counterparties $(i,j)\in V \times V$, the OIS contracts follow a marked point process \cite[Section 6.4]{daley2003introduction}

\begin{equation}
    \left\{ \left(t^{(i, j)}_k, N^{(i,j)}_k, K^{(i,j)}_k, T^{(i,j)}_k, \delta^{(i,j)}_k = (\delta^{(i)}_k, \delta^{(j)}_k) \right) \right\}_{k \in \Nb }
\label{eq:definition_contracts_as_point_process}
\end{equation}

where the ground point process $\{t^{(i, j)}\}_{k \in \Nb}$ models the start times of the contracts and the marks $\left(N^{(i,j)}_k, K^{(i,j)}_k, T^{(i,j)}_k, \delta^{(i,j)}_k = (\delta^{(i)}_k, \delta^{(j)}_k)\right)$ represent respectively the principal, fair OIS rate (as determined at time $t^{(i, j)}_k$ using Eq. \ref{eq:OIS_rate_definition}), the maturity, and the payer/receiver indicators for the $k$-th OIS contract. For simplicity, throughout this work, we assume that the counterparties can trade in one kind of standardized OIS, with constant principal equal to one and maturity equal to one year.

We adopt a reduced-form approach -- as commonly done, for instance, in credit risk \cite{giesecke_default_timings} when modelling default times -- and we assume that the point process $\{ t^{(i, j)}_k\}_{k \in \Nb}$ is a Cox process with stochastic intensity given by the following affine function of the reference rate

\begin{equation}
    \lambda_t^{(i,j)}\left(r(t), x_i, x_j\right) = \gamma\cdot \exp{\eta +\left(\theta+\beta\cdot g(x_i, x_j)\right)r(t)},
\label{eq:lambda_def_features}
\end{equation}

where $\gamma, \eta, \theta, \beta \in \Rb$ are parameters and the function $g(x_i,x_j)$ is a symmetric function that accounts for the features of each agent, defined as:

\begin{equation}
    g(x_i,x_j) = \frac{-x_ix_j + \mid x_i-x_j\mid + (x_i + x_j)}{3} = \begin{cases}
        +1 & \text{(hub-private)}\\1/3 & \text{(hub-hub)}\\-1 & \text{(private-private)}
    \end{cases}
\end{equation}

This network dynamic is aimed at capturing a basic stylized behavior of the financial entities: financial hubs trade frequently with privates and somewhat less frequently with other hubs, while privates are unlikely to enter an OTC derivative contract with other privates. Furthermore the contract generation process depends stochastically on the reference rate, which captures the prevailing market conditions.

In practice, since we work in a discrete-time framework, we simulate from a finite Cox process supported on the grid $\Tc$ by setting:

\begin{equation}
\label{eq:MPP_def}
t^{(i, j)}_k \leftarrow \inf \{ t \in \Tc \; : \; t^{(i,j)}_k \le t \}.
\end{equation}

The exact simulation procedure is discussed in detail in Appendix \ref{app:simulation}.

\paragraph{Variation margin forecasting problem.} We now formulate the actual forecasting problem we are interested in. Following Eq. \eqref{eq:OIS_value}, at time $t\in\Tc$, the mark-to-market value for counterpary $i$ of the $k$-th contract between the pair $(i,j)$ is:

\begin{equation}
V^{(i, j)}_k(t) = \delta^{(i)}_k N^{(i,j)}_k \left( p(t, T_k) \left( 1 + K^{(i,j)}_k(T_k - t_k) \right) - \prod_{t_i \in (t^{(i,j)}_k, t] \cap \Tc} \left( 1 + r(t_i) \Delta t \right)\right)\1_{\{ t \in (t_k, T^{(i,j)}_k]\}},
\label{Eq:k-th_value}
\end{equation}

where the value is equal to zero if the contract has not yet started or is already matured, as prescribed by the random indicator function on the right.

Therefore the net value of all outstanding contracts of node $i$ is simply the sum over all the contracts with its neighbors $\Nc_i(t)$ at time $t \in \Tc$:
\begin{equation}
    V^{(i)}(t) := \sum_{j\in\Nc_i(t)}{\sum_{k\in\Nb}{V^{(i,j)}_k(t)}}
\end{equation}

The net variation margin $M^{(i)}(t_{l+1})$ of node $i$ at time $t_{l+1}\in\Tc$ is simply the change in the mark-to-market value of $V^{(i)}$ from time $t_l$ to time $t_{l+1}$:
\begin{equation}
    M^{(i)}(t_{l+1}) := \sum_{\Nc_i(t)}{\sum_{k\in\Nb}{V^{(i,j)}_k(t_{l+1}) - (1+r(t_{l+1}) \Delta t)V^{(i,j)}_k(t_l)}},
\label{eq:variation_margin}
\end{equation}

where the mark-to-market value at time $t_l$ is capitalized by the overnight interest rate $r(t_{l+1})$ to ensure that $M^{(i)}(t_{l+1})$ is a martingale under $\Qb$.

To formalize the variation margin forecasting problem, we fix the following filtrations:
\begin{align}
    \Fc_t  &= \sigma\left(r(s), s\leq t \right) \\
    \Hc_t &= \sigma\left((t_k, N_k, K_k, T_k, \delta_k), \; \text{$k \in \Nb$ such that $t_k \le t$} \right)
\end{align}

\begin{figure}[h]
    \centering
    \includegraphics[width=1\textwidth]{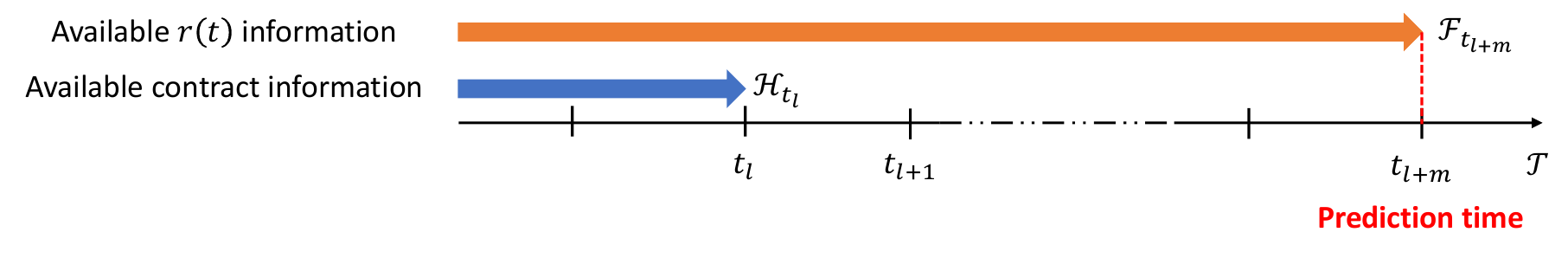}
    \caption{The $m$-steps ahead variation margin forecasting problem at time $t_l$.}
    \label{fig:prediction_problem}
\end{figure}

At time $t_l$ the goal is to predict the $m$-steps ahead net variation margin $M^{(i)}(t_{l+m})$ for all nodes in the graph. This is therefore an $m$-steps ahead node forecasting task. The forecast is done using the information available in $\Hc_{t_l}$, which is the information contained in all past and outstanding contracts, together with the information contained in $\Fc_{t_{l+m}}$, which includes the future evolution of the reference interest rate up to time $t_{l+m}$, as illustrated in Figure \ref{fig:prediction_problem}.

We include the information of the future reference rate trajectory because we want to obtain a conditional forecasting model, i.e. a model able to predict variation margins conditionally on a given scenario for the future evolution of the reference interest rate. Such a model, after being trained on real data, can be used as a stress testing tool to monitor the fragility of the OTC network to shocks in the reference rate.

Notice that the trajectory of the interest rate not only affects the value of already existing contracts, but also the future network dynamics. As the number $m$ of steps ahead increases, the model must increasingly include forecasted values for newly created contracts (or the absence thereof), thus being forced to learn the correct network dynamics.

The theoretical best predictor, which serves as a benchmark for our model, is the conditional expectation given by

\begin{equation}
    \prod_{t_i \in (t_l, t_{l+m}] \cap \Tc}{(1+r(t_i) \Delta t)^{-1}\EQ{M(t_{l+m})\mid \Fc_{t_{l+m}}\vee\Hc_{t_{l}}}}.
\label{eq:benchmark}
\end{equation}

\section{Proposed DGNN model}
\label{sec:model}
Solving the $m$-steps ahead variation margin forecasting problem requires aggregating information from the counterparty's neighbors, as these node and edge features contain information about the network dynamics.

\begin{figure}[h]
    \centering
    \includegraphics[width=1\textwidth]{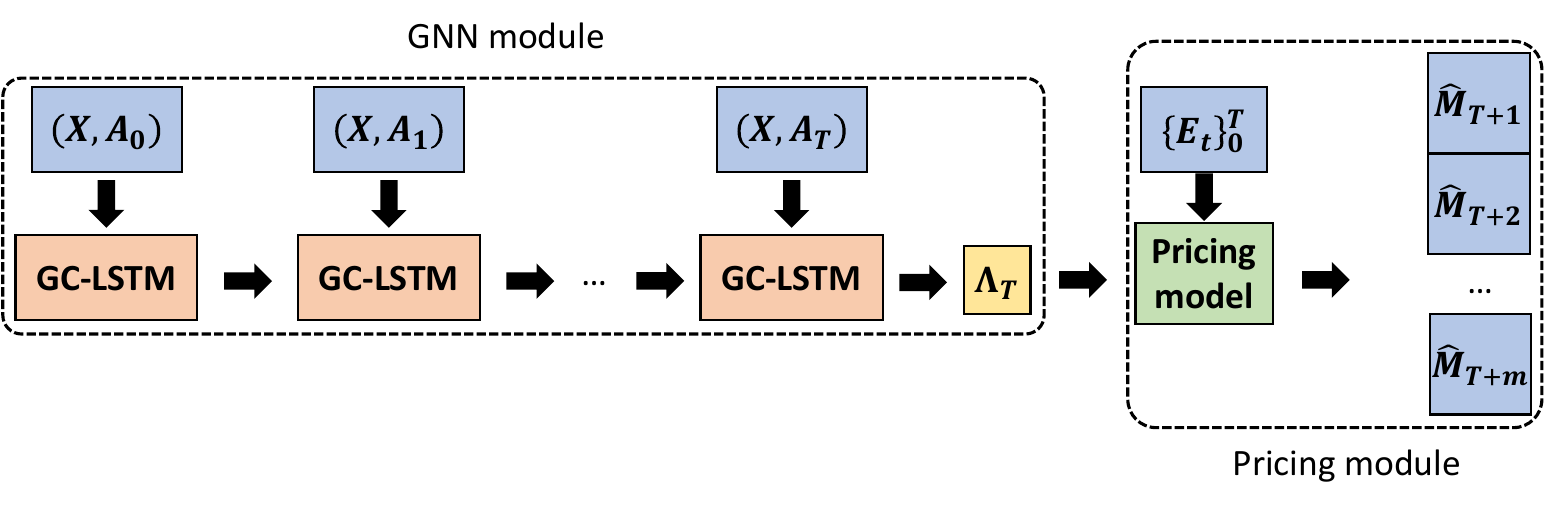}
    \caption{Architecture of the proposed DGNN model.}
    \label{fig:complete_architecture}
\end{figure}

We propose a Dynamic Graph Neural Network (DGNN) model capable of taking into account both the topology of the graph and its dynamic evolution. The devised architecture consists of two main components: a \emph{GNN module} and a \emph{pricing module}, which we introduce below. Both modules are recurrent, thus capturing the temporal evolution of the input data. The GNN module receives a sequence of snapshots of the temporal graph and endeavors to learn a latent node representation of the stochastic intensities $\lambda^{(i,j)}_t$, while the pricing module is trained on contract data and aims to learn a latent representation suitable for pricing. An illustration of the architecture is shown in Figure \ref{fig:complete_architecture}.

\subsection{GNN module} 

The GNN module consists of a GC-LSTM \cite{gc-LSTM}, a dynamic Graph Neural Network (DGNN), which receives snapshots of the temporal network. Its objective is to produce latent embeddings able to capture the stochastic intensity governing the appearance of new edges. As clear from Equation \ref{eq:lambda_def_features} the stochastic intensity $\lambda^{(i,j)}_t$ depends on the node features of the involved agents and the prevailing market conditions represented by $r(t)$. 

The purpose of this module is to provide a representation of node $i$'s features along with those of its typical neighbors, thereby aggregating neighbor features, regardless of the market conditions. This is achieved through a neighborhood aggregation scheme implemented by this GNN architecture. At each time step, the model receives the node feature matrix $X$ along with the undirected adjacency matrix $A_t$, with entries in $\{0,1\}$, which encodes only the presence or absence of an outstanding contract between the two counterparties. After processing the training window (for a detailed description of the training process, refer to Appendix \ref{app:training}), the model generates an embedding $\Lambda_T \in \Rb^{N \times h_\lambda}$, where $h_\lambda$ is the hidden size of this embedding and is a tunable hyper-parameter.

\begin{figure}[h]
    \centering
    \includegraphics[width=0.9\textwidth]{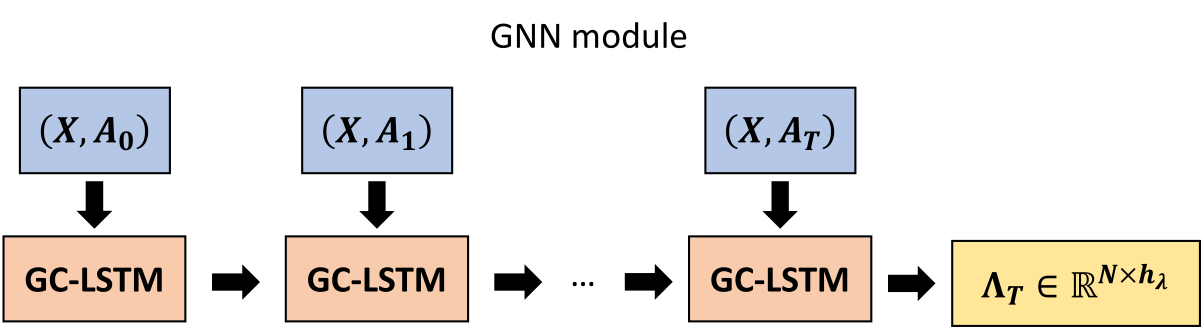}
    \caption{GNN module.}
    \label{fig:gnn_module}
\end{figure}

\subsection{Pricing module}

The pricing module is designed to generate the vector of forecasted variation margins $\hat{M}^{(i)}(t)$ based on the already existing contracts and the expected newly created contracts in the time interval $(t_l, r_{l+m}]$. To simplify the exposition we first explain how the $1$-step ahead forecasts are created and we then move on to the $m$-steps ahead case.

\paragraph{$1$-step ahead.} The pricing module needs to learn the relationship between the input contract features and the variation margin. Each contract at time $t$ is represented by the following features:

\begin{equation}
    \left[(T-t)/365, p(t_0,T), p(t,T), B(t_0), B(t), \delta\right],
\label{eq:contract_quantities}
\end{equation}

which are observable on the market at time $t$. In particular $p(t, T)$ denotes the price of a bond at time $t$ with maturity $T$, as introduced in Eq. \eqref{eq:bond_price_def}, and $B(t)$ is simply the value of the risk-free asset capitalized at the reference rate, i.e.

\begin{equation}
    B(t)\coloneqq\prod_{t_i \in (0, t] \cap \Tc}{\left(1+r(t_i) \Delta t\right)}
\label{eq:B_def}
\end{equation}
\begin{figure}[h]
    \centering
    \includegraphics[width=0.9\textwidth]{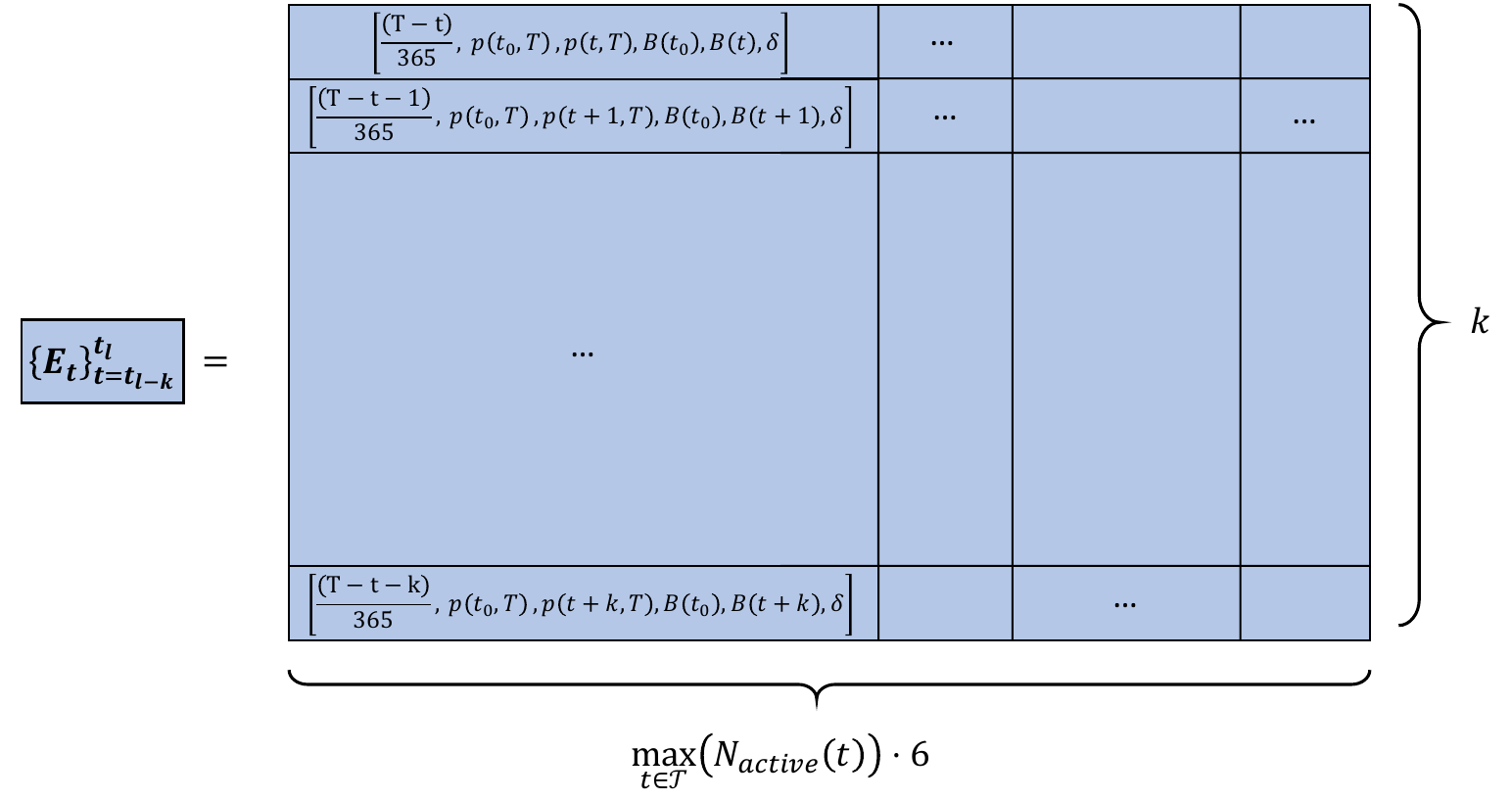}
    \caption{Contract matrix passed to the pricing module in the scenario involving two nodes.}
    \label{fig:contract_matrix}
\end{figure}

The contract features for each contract in the last $k$ steps (where $k$ is the length of the training window) are represented jointly in a \emph{contract matrix}, as depicted in Figure \ref{fig:contract_matrix}. Since the number of outstanding contracts is variable, we set the width of the contract matrix to the maximum number of simultaneously active contracts observed over the entire time grid $\Tc$, multiplied by the number of contract features.

\begin{figure}[h]
    \centering
    \includegraphics[width=0.6\textwidth]{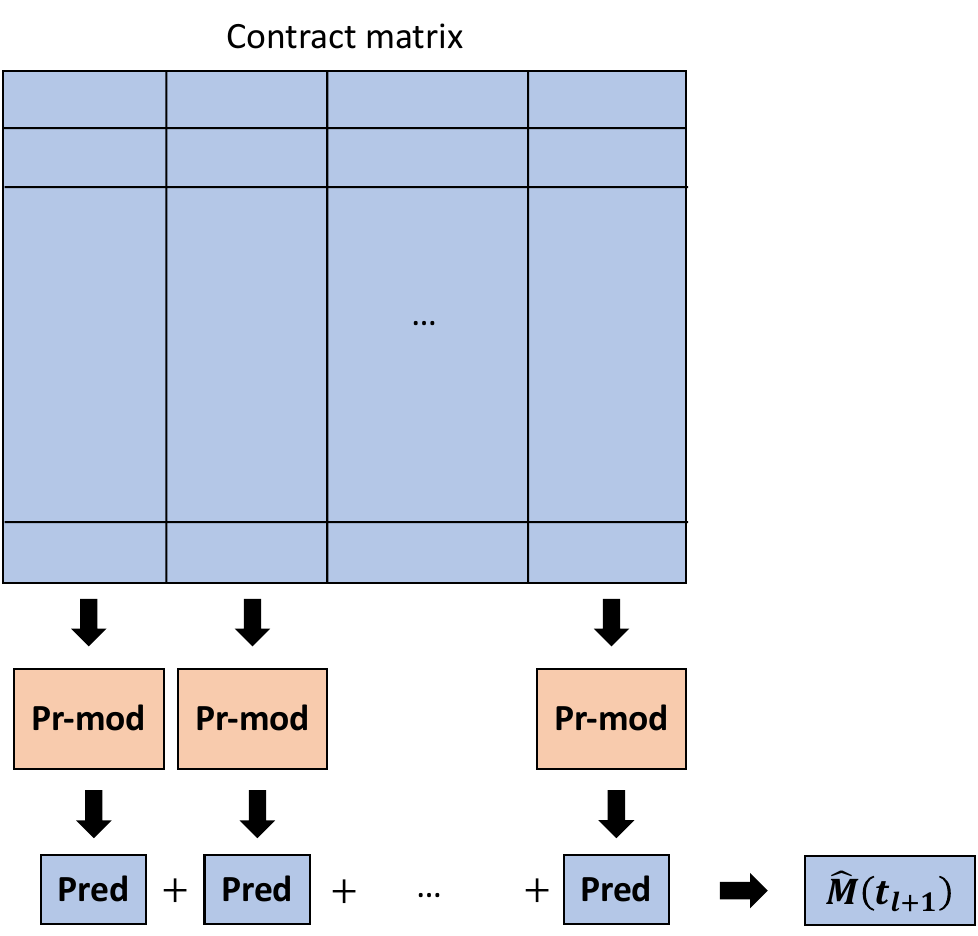}
    \caption{Architecture of the pricing module, consisting of parallel applications of the Pr-mod module.}
    \label{fig:1_step_ahead_pricing_module}
\end{figure}

The pricing module consists of several parallel applications of a single \emph{Pr-mod} module that takes as input every non-zero column of the contract matrix and outputs a forecasted variation margin prediction. These single-contract variation margin predictions are then summed up to produce the final net variation margin prediction $\hat{M}(t_{l+1})$, as depicted in Figure \ref{fig:1_step_ahead_pricing_module}.

\begin{figure}[h]
    \centering
    \includegraphics[width=0.9\textwidth]{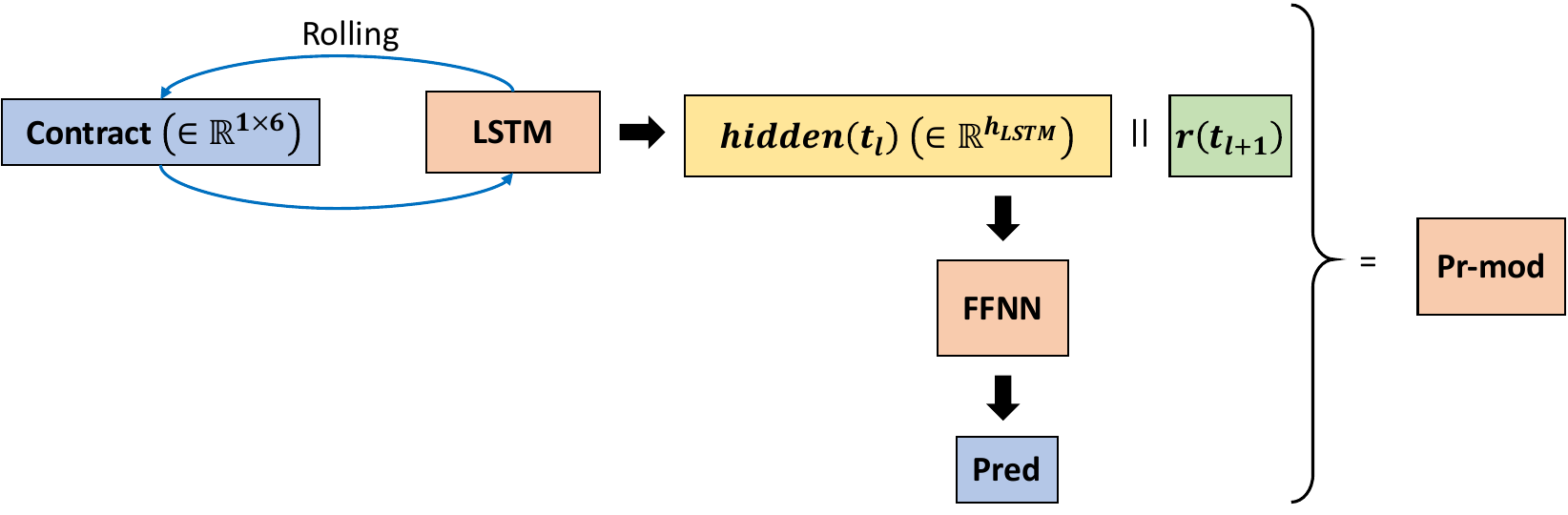}
    \caption{Architecture of the 'Pr-mod' module.}
    \label{fig:pr_mod}
\end{figure}

The Pr-mod module itself is a combination of an LSTM and a multi-layer Feed-Forward Neural Network (FFNN), as illustrated in Figure \ref{fig:pr_mod}. It is designed to learn the pricing of individual contracts, taking as input a single column of the contract matrix. The LSTM recurrently generates a latent representation of the contract at time step $t_l$ of fixed size $h_{LSTM}$, which is concatenated with the value of the interest rate $r(t_{l+1})$ at time $t_{l+1}$, acting as the conditioning variable in our conditional forecasting problem. Subsequently, this vector of size $h_{LSTM} + 1$ is fed into the FFNN, producing a real-valued prediction of the variation margin associated to a single contract in the contract matrix.

\paragraph{$m$-steps ahead.} The $m$-steps ahead forecast relies on the Sequence-to-Sequence framework. At each time step, the model generates a prediction of the contract matrix shifted one time step in the future by deleting the row corresponding to the $t_{l-k}$ time step and predicting a new row for the $t_{l+1}$ time step.

\begin{figure}[h]
    \centering
    \includegraphics[width=0.9\textwidth]{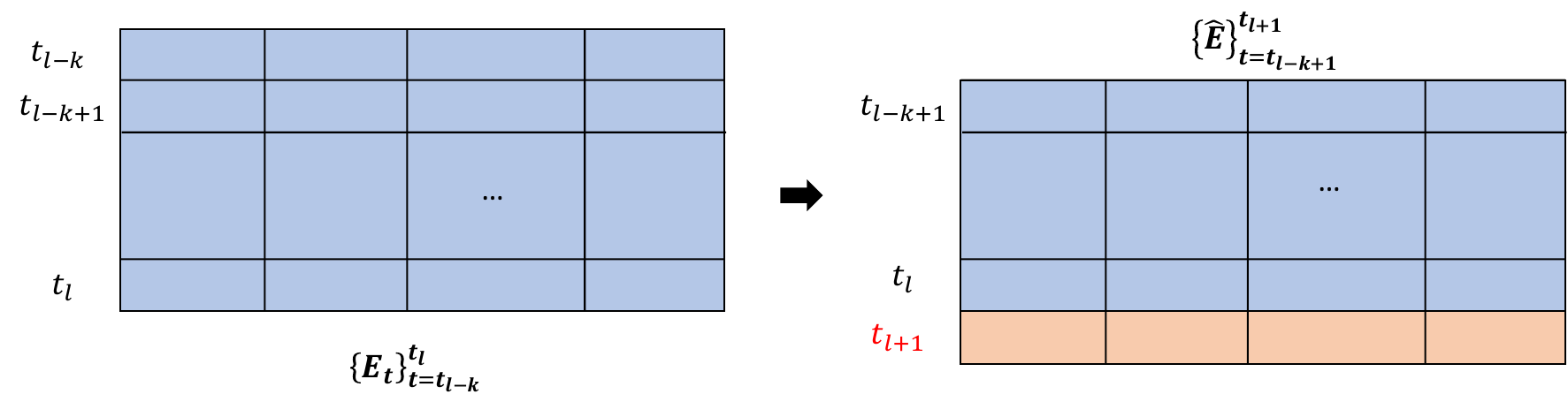}
    \caption{Contract prediction process.}
    \label{fig:contract_extrapolation}
\end{figure}

In other words, the contract matrix $\{E\}_{t=t_{l-k}}^{t_l}$ is replaced by a new contract matrix $\{\hat{E}\}_{t=t_{l-k+1}}^{t_{l+1}}$ by deleting the first row and appending a new row, as illustrated in Figure \ref{fig:contract_extrapolation}.

\begin{figure}[h]
    \centering
    \includegraphics[width=0.9\textwidth]{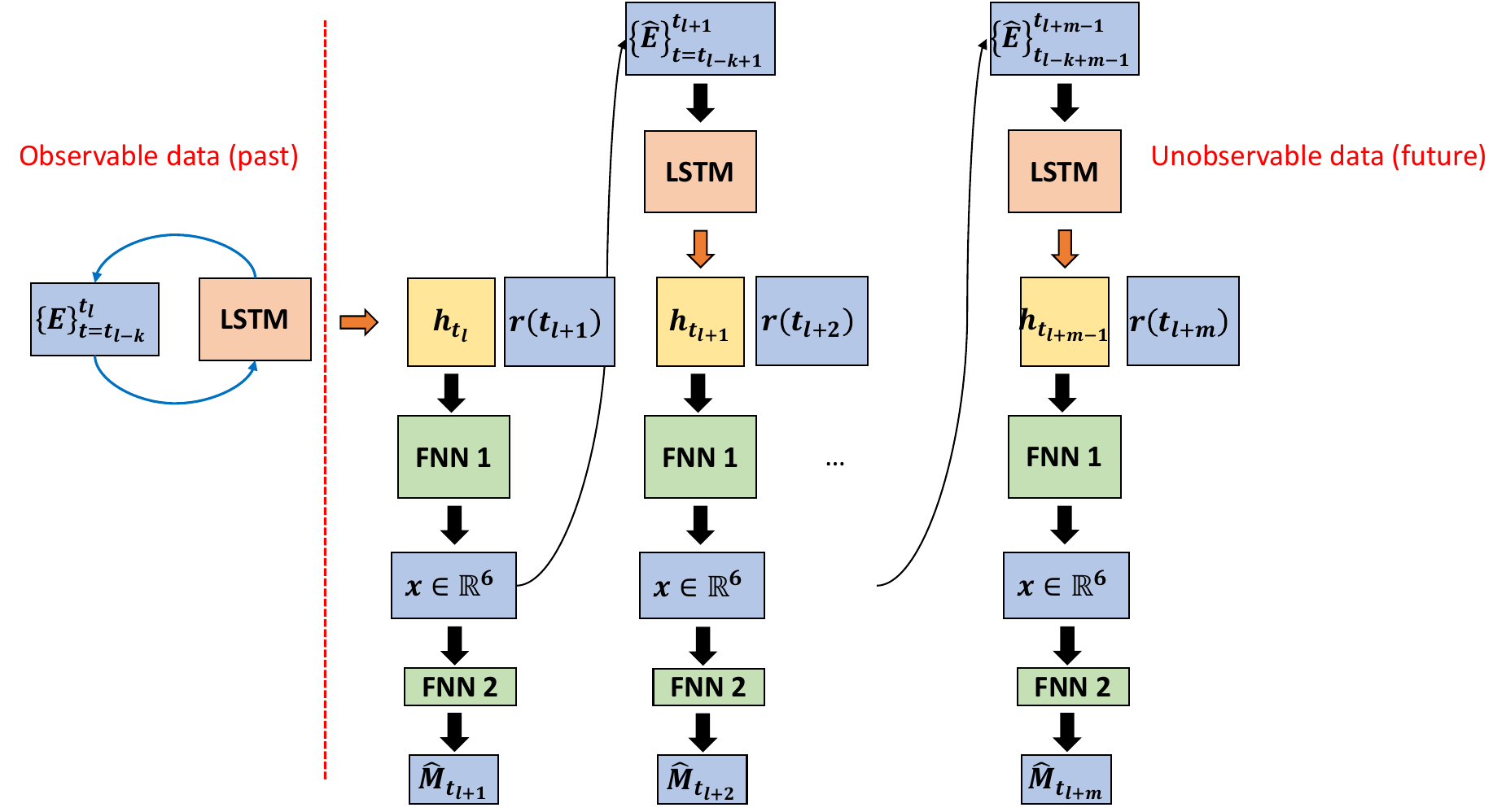}
    \caption{Multi-step ahead forecasting process.}
    \label{fig:1_node_multi_step}
\end{figure}

This process is iteratively repeated for each one of the $m$ steps in the future, until the prediction's time $t_{l+m}$. Each of the extrapolated contract matrices is used similarly to the 1-step ahead scenario to produce a prediction $\hat{M}_{t_{l+i}}$, for $i=1, \ldots, m$. Despite the primary focus being on time step $t_{l+m}$, the model is trained by computing the loss on each time step's prediction $\{\hat{M}_{t_{l+i}}\}_{i=1}^{m}$. The full multi-step ahead forecasting process is depicted in Figure \ref{fig:1_node_multi_step}.

This iterative approach enables the model to learn from predicting the variation margin over multiple time steps, improving the performance at the final forecasting time $t_{l+m}$.

\begin{figure}[H]
    \centering
    \includegraphics[width=0.9\textwidth]{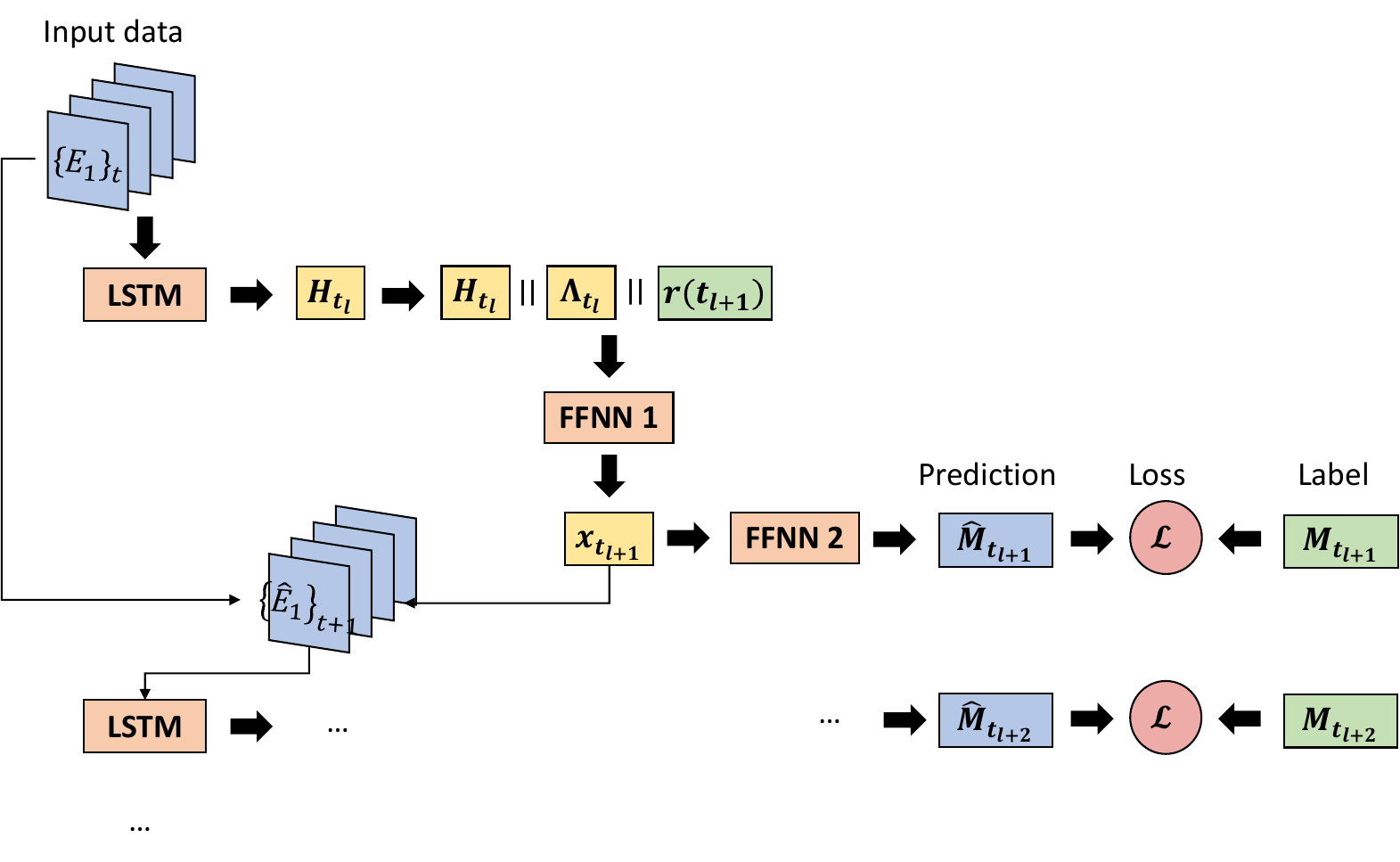}
    \caption{Complete pricing module architecture, including node embeddings from the GNN module.}
    \label{fig:complete_pricing_module}
\end{figure}

\paragraph{Adding the latent node embedding.} At this point we still need to discuss how the node embeddings produced by the GNN module are actually used as inputs by the pricing module. As shown in Figure \ref{fig:complete_pricing_module}, which shows the complete pricing module architecture, this is done simply by concatenation. More specifically, the pricing module is supplied with a contract matrix which is processed recursively by the LSTM, thus producing latent representations $H_{t_l}\in\Rb^{h_{\text{LSTM}}}$. These latent representations are then concatenated with the stochastic interest rate $r(t_{l+1})$ and the node embeddings $\Lambda_{t_l}$ from the GNN module, thus capturing the stochastic intensity of the edge process. 

The resulting matrix is then fed into two FFNNs: FFNN1, which generates the next-step predicted contract features $x_{t_{l+1}}\in\Rb^{N\times6}$ used to produce the next-step predicted contract matrices, and FFNN2, which outputs the predicted variation margin vector $\hat{M}_{t_{l+1}}\in\Rb^{N}$ exactly as in Figure \ref{fig:1_node_multi_step}.

\section{Experiments}
\label{sec:experiments}
We now present experimental results for the $m$-steps ahead variation margin forecasting problem using our DGNN model, ranging from the $1$-step ahead prediction for a single node, to the complete $N$-nodes scenario.

\begin{figure}[H]
    \centering
    \includegraphics[width=1\textwidth]{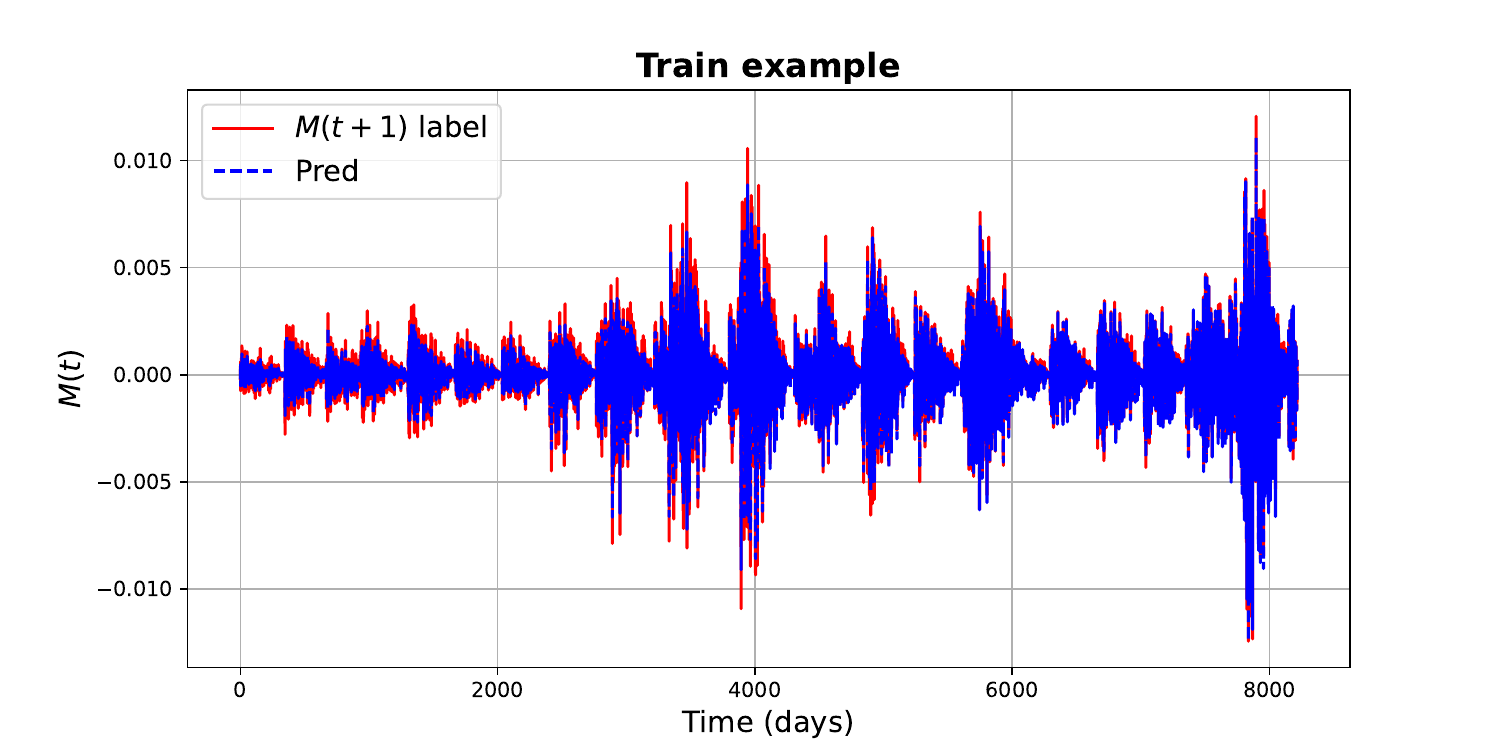}
    \caption{Single-step ahead variation margin forecast on the training set.}
    \label{fig:1_node_1_step_train_margin_lambda}
\end{figure}

\begin{figure}[H]
    \centering
    \includegraphics[width=1\textwidth]{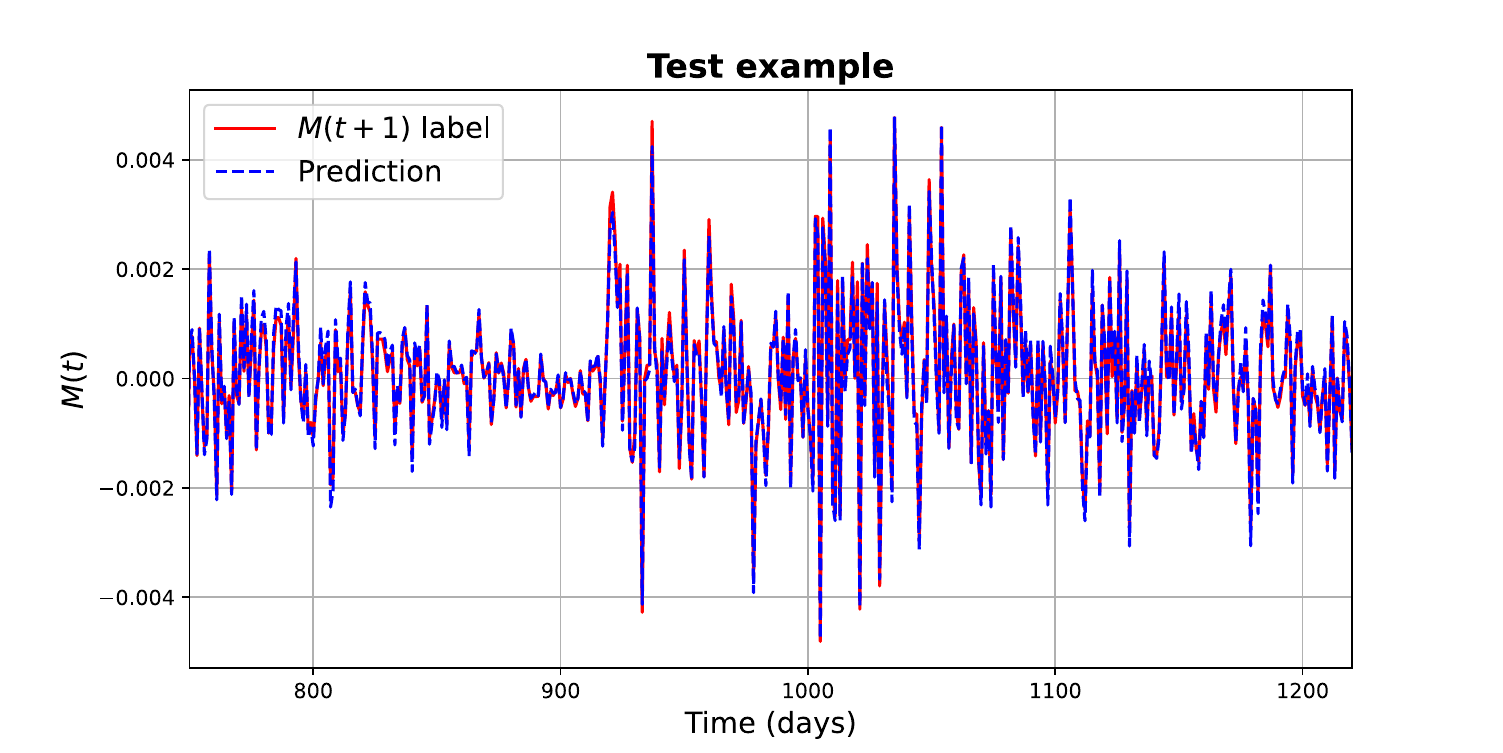}
    \caption{Single-step ahead variation margin forecast on the test set.}
    \label{fig:1_node_1_step_test_margin_lambda}
\end{figure}

\paragraph{$1$-step ahead} Results for the training and test sets are depicted in Figure \ref{fig:1_node_1_step_train_margin_lambda} and in Figure \ref{fig:1_node_1_step_test_margin_lambda}, respectively.

The model operates on individual time windows obtained from a standard windowing procedure, as discussed in Appendix \ref{app:training}. The plots presented here are obtained by concatenating the forecasts of consecutive windows into a sequence, where each point represents a single forecast made from a window, allowing the creation of the continuous line that can be observed in the plots.

As clear from Figure \ref{fig:1_node_1_step_train_margin_lambda}, the simulation dataset is rich enough to allow the model to learn complex dependencies in different regimes. Figure \ref{fig:1_node_1_step_test_margin_lambda} demonstrates that the model is able to generalize well to a test dataset. The encouraging results underscore the model's ability to learn pricing as well as the temporal processes of contracts, despite the simplicity of the 1-step ahead example.

\begin{figure}[H]
    \centering
    \includegraphics[width=1\textwidth]{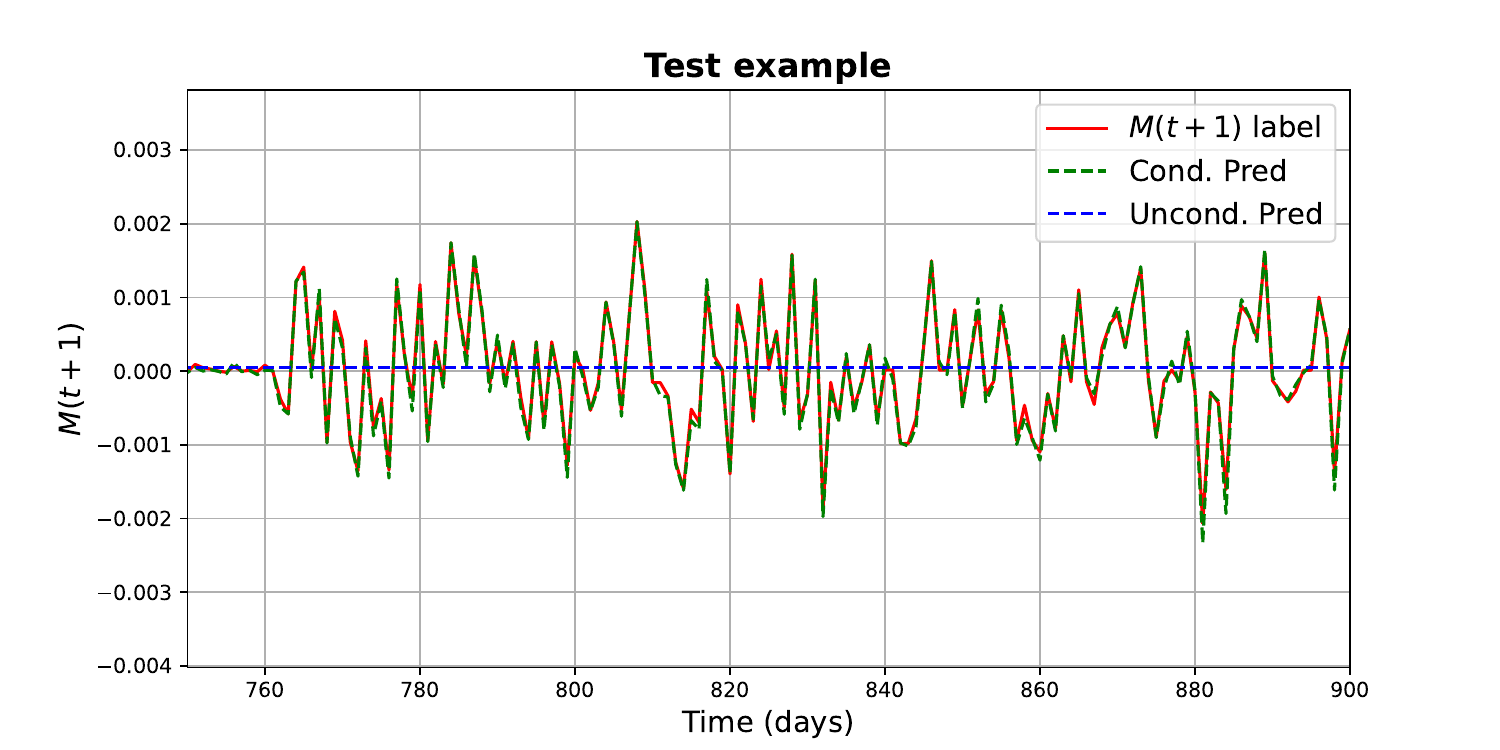}
    \caption{Model predictions on the test set with and without conditioning input.}
    \label{fig:1_step_1_node_test_margin_nocond_cond}
\end{figure}

The conditioning information provided by the interest rate $r(t_{l+1})$ is crucial in the forecasting problem. To test whether the model is effectively leveraging this additional piece of information, passed through a simple concatenation, we compare the model performance with and without the conditioning input. The results for an interval of the test set are presented in Figure \ref{fig:1_step_1_node_test_margin_nocond_cond}. The model fails to generalize in the non-conditioned case, as it can only learn the zero expected value of the variation margins, while the conditioned prediction perfectly matches the test data.

\paragraph{$m$-steps ahead}

For the $m$-steps ahead prediction, it is important to evaluate the model's performance against the theoretical best predictor introduced in Equation \eqref{eq:benchmark}. As this quantity relies on the expectation of the process governing the arrival of contracts, it is necessary to use Monte Carlo methods to compute this expectation. 

\begin{figure}[H]
    \centering
    \includegraphics[width=1\textwidth]{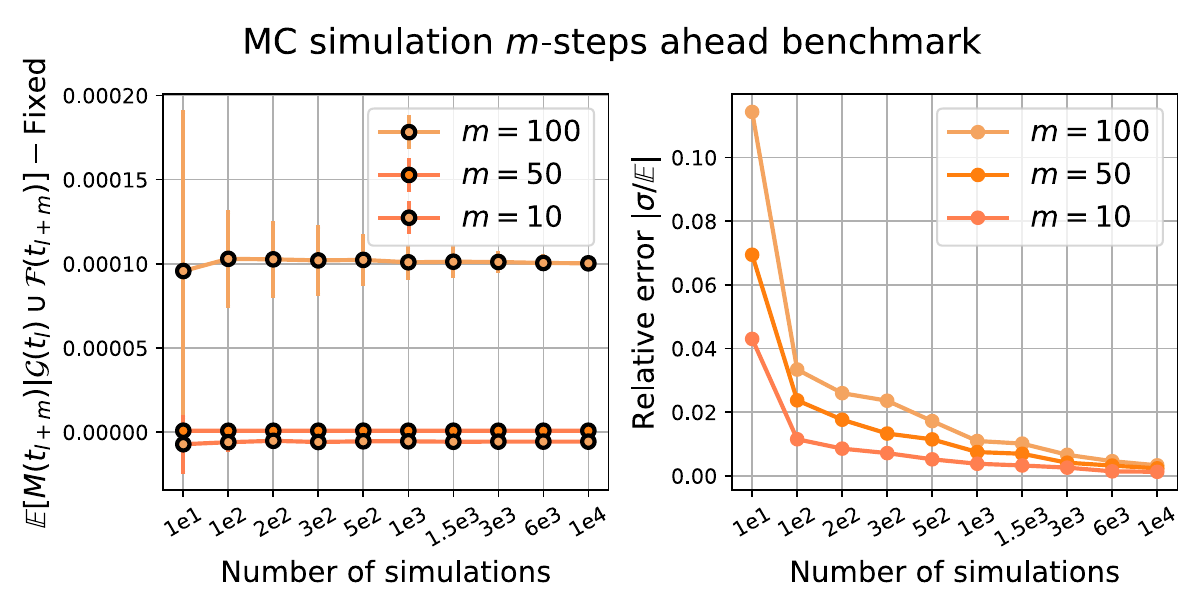}
    \caption{Number of simulations needed to compute the benchmark within given accuracy.}
    \label{fig:MC_simulation_benchmark}
\end{figure}

More specifically, this benchmark comprises two components: the expectation of the variation margin for contracts existing at time $t_l$ and remaining active until time $t_{l+m}$, which is known in advance given the $m$-steps interest rate trajectory (it is $\Fc_{t_l}$-measurable), and the expectation of the variation margin from contracts that may begin in the interval $[t_{l+1}, t_{l+m-1}]$. This second term is more complex and it becomes the dominant term as the number of steps ahead increases. As depicted in the right plot of Figure \ref{fig:MC_simulation_benchmark}, achieving a relative error below $2\%$ in the benchmark dispersion with $m\leq21$ days (3 weeks) requires approximately $10^3$ simulations of the contract arriving process in the interval $[t_{l+1}, t_{l+m-1}]$.

The left plot displays the difference between the benchmark and its fixed component, derived from contracts existing at time $t_l$ and remaining active up to time $t_{l+m}$, for different values of $m$, with the number of simulations used for the Monte Carlo expectation varying. The plot was obtained repeating the computation $10^2$ times. The right plot shows the relative error for the benchmark across different values of $m$, with the number of simulations in the Monte Carlo expectation also varying. This plot was also obtained repeating the computation $10^2$ times.

\begin{figure}[H]
    \centering
    \includegraphics[width=1\textwidth]{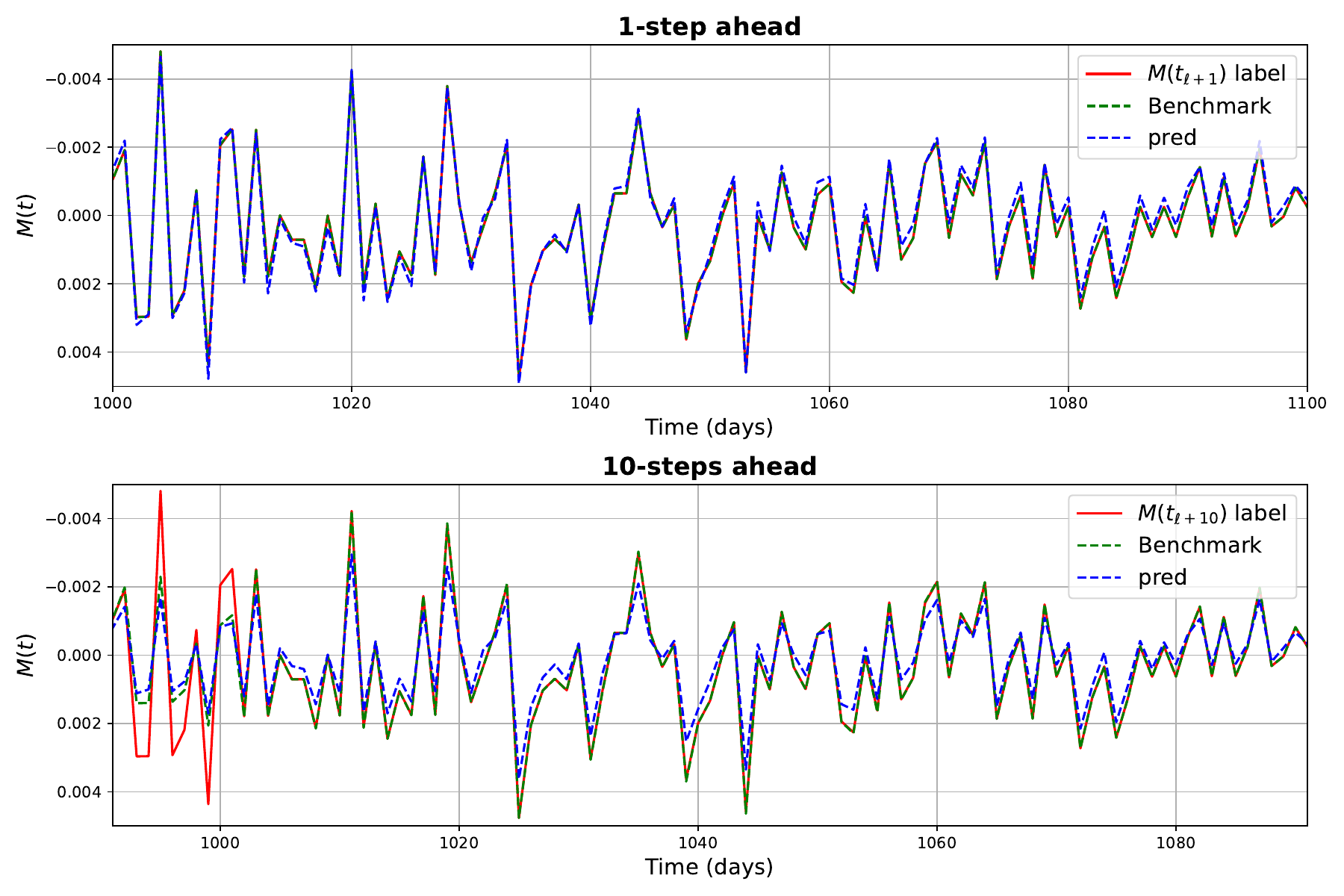}
    \caption{Comparison between a single-step ahead prediction and a $10$-steps ahead forecast of the variation margin.}
    \label{fig:1_step_10_steps}
\end{figure}

Once the benchmark is established, the model's performance in the $m$-steps ahead prediction can be assessed. Figure \ref{fig:1_step_10_steps} illustrates the performance comparison between a single-step ahead forecast and a $10$-steps ahead forecast.

Notably, in the single-step ahead scenario, the benchmark perfectly aligns with the labels since the second term of the expectation relative to the contract arrival is zero. This occurs because contracts arriving at $t_{\ell+1}$ do not affect $M(t_{\ell+1})$.

Conversely, in the $10$-steps ahead scenario, labels consistently deviate from the benchmark. The deviation arises because the labels represent a realization of the contract arrival process, while the benchmark reflects the conditional expectation of such a process.

\begin{figure}[h]
    \centering
    \includegraphics[width=1\textwidth]{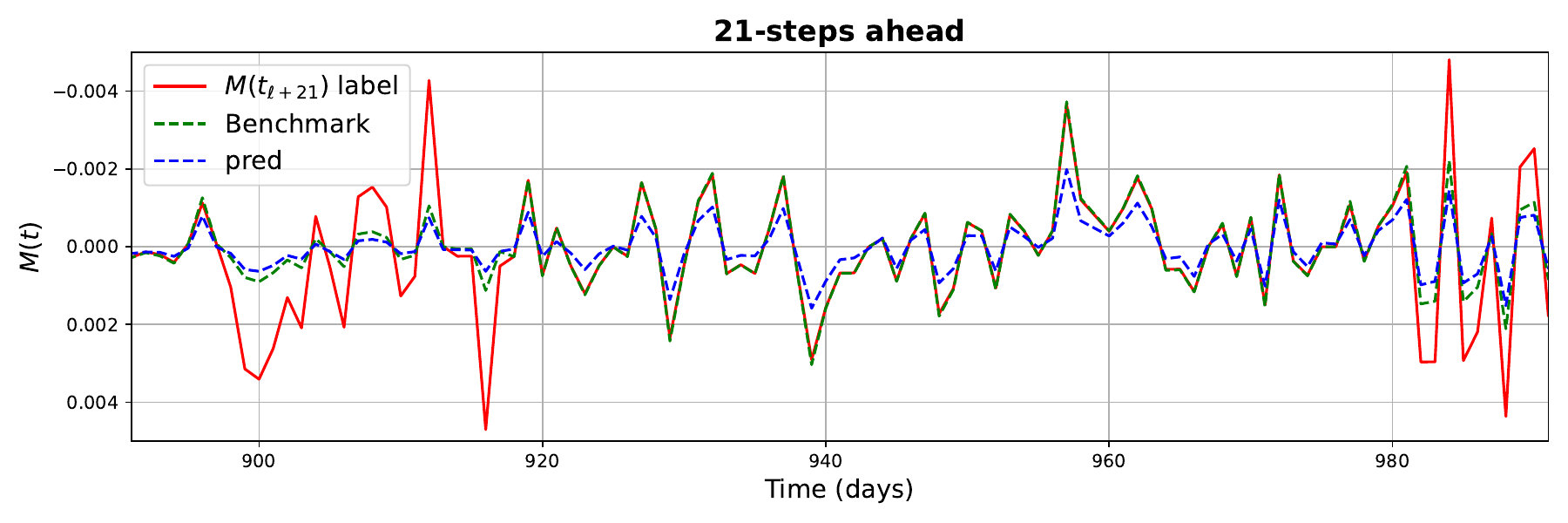}
    \caption{21-steps ahead variation margin forecast on the test set.}
    \label{fig:21_steps_ahead}
\end{figure}

Figure \ref{fig:21_steps_ahead} depicts model's performance in the $21$-steps ahead scenario, the most extensive forecasting horizon tested here.

Even in this scenario, the model generalizes effectively, with its predictions closely resembling the theoretical best predictor. The $21$-steps ahead prediction represents the furthest forecasting attempted.

\begin{figure}[h]
    \centering
    \includegraphics[width=0.9\textwidth]{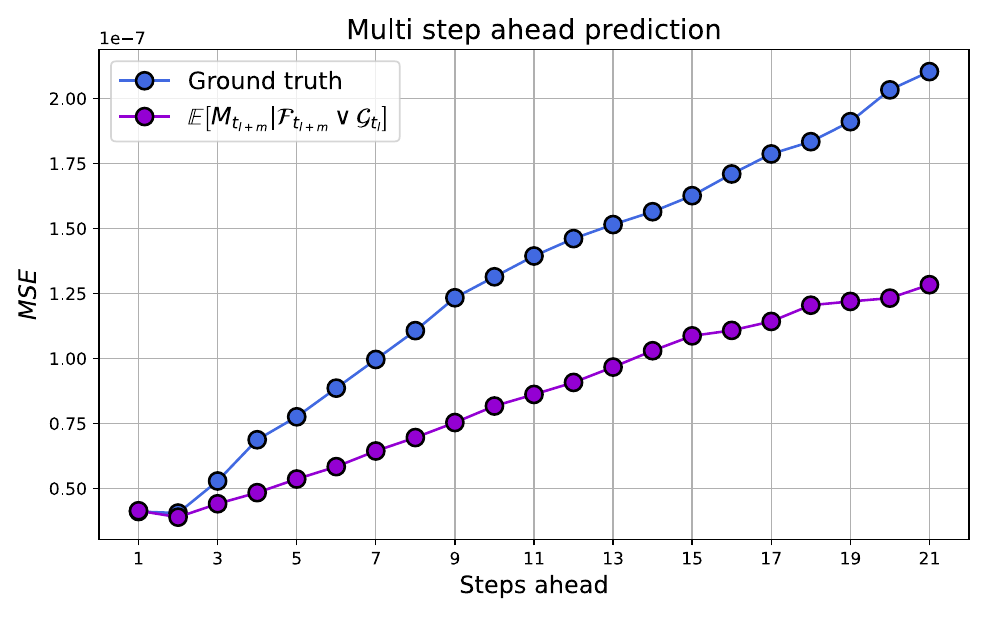}
    \caption{MSE loss in the multi-step ahead forecasting task, with varying numbers of steps ahead, computed with respected to the theoretical best predictor (benchmark) and the true labels (ground truth).}
    \label{fig:benchmark_vs_labels}
\end{figure}

Figure \ref{fig:benchmark_vs_labels} show the Mean Squared Error (MSE) loss between the predictions and the benchmark, as well as the Mean Squared Error (MSE) loss between the predictions and the test labels, across varying numbers of the steps ahead in the forecasting task. Clearly, in both scenarios, the mean squared error increases as the number of steps in the future increases, reflecting the model's decreasing access to information for its forecasting task. However, it is evident that predictions exhibit better alignment with the benchmark, once again highlighting the distinction between learning a realization of the contract process, here represented by the labels, and the expectation of it, embodied by the benchmark.

The full model and training specifics used to obtained the previous results can be found in Appendix \ref{app:training}.

The process outlined is computationally intensive, limiting the ability to analyze networks with a large number of nodes. The results presented here are therefore preliminary, demonstrating that the proposed approach is effective but highlighting the need for further development to enhance scalability. 

\begin{figure}[h!]
    \centering
    \includegraphics[width=0.9\textwidth]{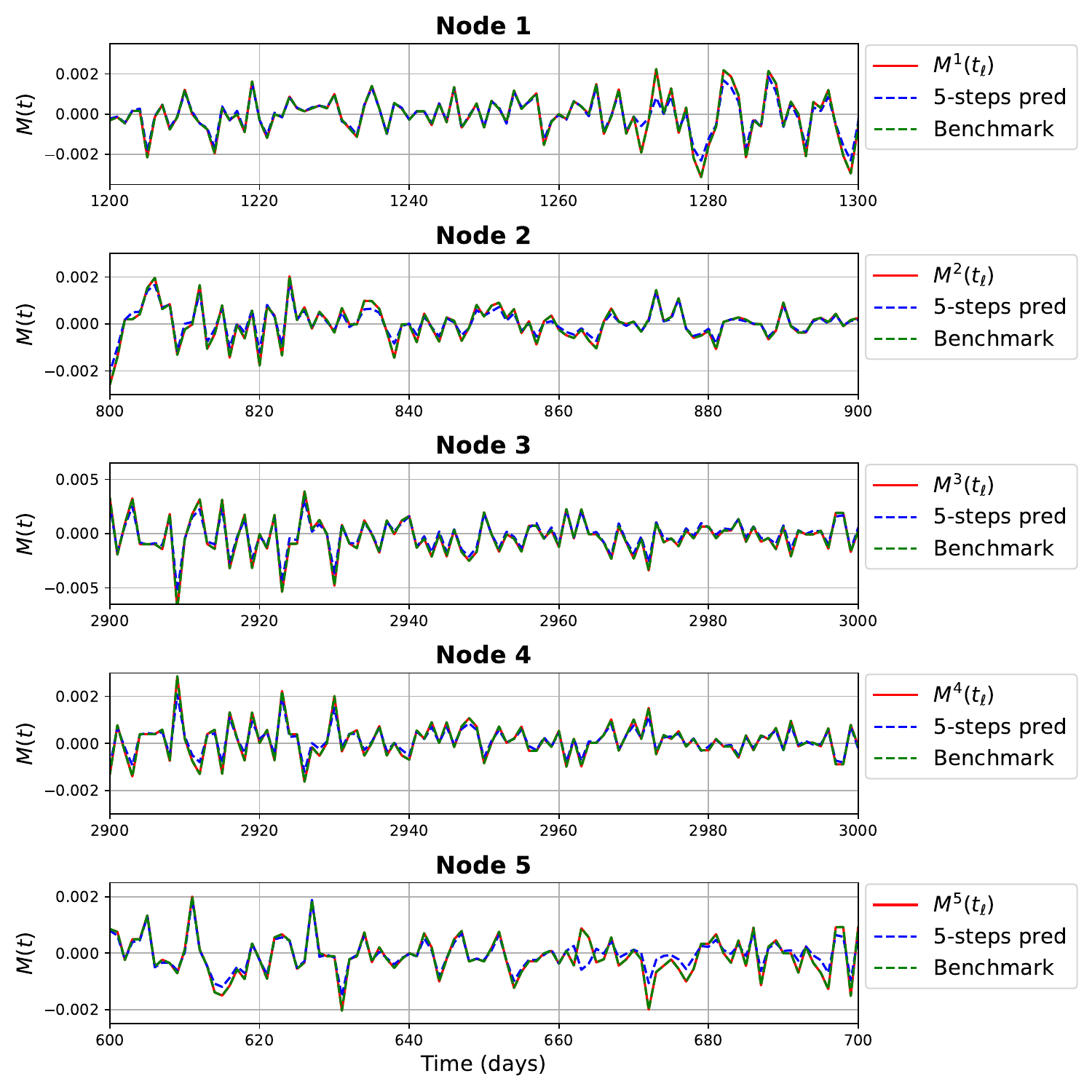}
    \caption{5-steps ahead variation margin forecast on the test set for a network with $5$ nodes.}
    \label{fig:3_nodes_2_steps}
\end{figure}

Figure \ref{fig:3_nodes_2_steps} showcases the model's predictions for a network with $5$ nodes in a $5$-steps ahead scenario. The network is simulated over a $60$-year span, and detailed contract process parameters are provided in Appendix \ref{app:training} together with the training process specifics.

While this result is preliminary and lacks fine-tuning and other performance enhancing techniques, it serves as an initial confirmation that Dynamic Graph Neural Networks, combined with RNNs, can effectively  generalize out-of-sample from a training financial network dataset. 

\section{Conclusion}
\label{sec:conclusion}

We introduced a novel Dynamic Graph Neural Network (DGNN) model for solving a conditional $m$-steps ahead forecasting problem in a dynamic financial network. We validated our proposed DGNN on simulated data capturing stylized features of Overnight Indexed Swap (OIS) networks, in which financial entities trade OIS contracts dynamically and the network topology evolves conditionally on the OIS reference rate. 

The proposed model, consisting of a GNN module and a pricing LSTM module, is able to capture the evolving structural features of the network and generalize to unseen data. By effectively leveraging additional information from a conditioning variable, such as the OIS reference rate, the model can be used to produce accurate forecasts up to a $21$-day horizon under pre-determined stress test scenarios.

This proof-of-concept work shows that information on the trading network structure can be successfully incorporated into stress-testing practices, thus producing forecasts that take into account the dynamic and interconnected nature of the financial system.

By being trained on real historical data instead of relying on simulation studies, this method could offer valuable insights into how the market reacts to shocks in reference interest rates and other macro-economic variables, providing regulators and policymakers with a crucial tool for systemic risk monitoring.


\bibliographystyle{plain} 
\bibliography{refs} 

\appendix

\section{Simulation details}
\label{app:simulation}

\subsection{Interest rate process}
\label{subsec:interest_rate_process}

As clear from preceding sections, the reference spot interest rate plays a crucial role in shaping a market model; therefore modelling a stochastic process that characterizes the rate dynamics is pivotal, as it drives many mathematical properties of the market itself. Without delving too deeply in the field, one-factor short-rate models have a rich history \cite{brigo_mercurio} because directly modeling such dynamics is highly convenient. As a matter or facts, all fundamental quantities, such as rates and bonds, can be readily defined by no-arbitrage arguments, as the expectation of a functional of the process $r$. 

For instance, assuming the existence of a risk-neutral measure in a filtered probability space $(\Omega, \{\Fc_t\}_{t\geq0}, \Qb)$ implies that the arbitrage-free price at time $t$ of a contingent claim with payoff $H_T$ at time $T$ is given by:
\begin{equation}
    H_t = \EQ{D(t,T)H_t\mid \Fc_t} = \EQ{e^{-\int_t^T{r(s)ds}}H_T \mid \Fc_t}
\end{equation}
where the expectation is conditioned on the available information up to time $t$, here represented by the canonical filtration $\Fc_t$ to which $r$ is adapted. Specifically, the zero-coupon-price at time $t$ for maturity $T$ is characterized by a unit amount of currency available at time $T$, i.e. $H_T=1$, resulting in:
\begin{equation}
    p(t,T) = \EQ{e^{-\int_t^T{r(s)ds}} \mid \Fc_t}
\label{eq:general_bond_price}
\end{equation}
From this equation, it becomes evident that whenever it is possible to characterize the distribution of $e^{-\int_t^T{r(s)ds}}$ in terms of a chosen dynamics of $r$, conditional on the information available up to time $t$, it is possible to compute bond prices $p$ \cite{brigo_mercurio}. 

In this work, the focus is narrowed to the approach proposed by Cox, Ingersoll and Ross in 1985 \cite{CIR_model}. The model they introduced entails a "square-root" term in the diffusion coefficient of the instantaneous short-rate dynamics $r(t)$ and has served as a benchmark for many years due to its analytical tractability in the continuous case, the fact that the instantaneous rate is always positive, and the possibility of exact simulation \cite{glasserman}. 
\begin{definition}[CIR model]
Under the risk neutral measure $\Qb$, the CIR model is governed by the following differential equation:
\begin{equation}
    \begin{cases}
            dr(t) = k(\theta - r(t))dt + \sigma\sqrt{r(t)}dW(t)\\
            r(0)=r_0
        \end{cases}
\label{eq:CIR_definition}
\end{equation}
        
\end{definition}
Where $dW(t)$ denotes a $\Qb$-Wiener process and $k$, $r_0$, $\theta$, $\sigma$ are positive constants. The condition:
\begin{equation}
    2k\theta > \sigma^2
\end{equation}
ensures that the origin remains inaccessible to the process so that $r(t)$ remains positive $\Qb$ $a.s.$. The intuitive idea of Eq. \eqref{eq:CIR_definition}, is that $r(t)$ is pulled towards $\theta$ at a speed controlled by $k$.

Eq. \eqref{eq:CIR_definition} is not explicitly solvable; however, the transition density for the process is known \cite{feller}. Notably, the distribution of $r(t)$ given $r(u)$ for some $u<t$ is, up to a scale factor, a \textit{noncentral}-$\chi^2$ distribution. A noncentral $\chi^2$ random variable ${\chi'}_\nu^2(\lambda)$ with $\nu$ degrees of freedom, and noncentrality parameter $\lambda$, has a distribution given by:
\begin{align}
    \Pb\left({\chi'}_\nu^2(\lambda) \leq y\right) &= F_{{\chi'}_\nu^2(\lambda)}(y)\\
    &=e^{-\lambda/2}\sum_{j=0}^{\infty}{\frac{(\frac{1}{2}\lambda)^j/j!}{2^{(\nu/2)+j}\Gamma(\nu/2+j)}\int_0^y {z^{(\nu/2)+j-1}e^{-z/2}dz}}
\end{align}
for any $y>0$. Therefore the transition law for $r(t)$ in Eq. \eqref{eq:CIR_definition} given $r(u)$, can be expressed as:
\begin{equation}
    r(t) = \frac{\sigma^2 (1-e^{-k(t-u)})}{4k}{\chi'}_d^2\left(\frac{4ke^{-k(t-u)}}{\sigma^2 (1-e^{-k(t-u)})}r(u)\right), \quad t>u
\label{eq:CIR_transition}
\end{equation}
Where the degrees of freedom $d$ are given by:
\begin{equation}
    d = \frac{4\theta k}{\sigma^2}
\label{eq:d_dof}
\end{equation}
Eq. \eqref{eq:CIR_transition} essentially states that given $r(u)$, $r(t)$ is distributed as $\sigma^2 (1-e^{-k(t-u)})/4k$ times a noncentral-$\chi^2$ random variable with $d$ degrees of freedom given by Eq. \eqref{eq:d_dof} and non-centrality parameter $\lambda$ given by:
\begin{equation}
    \lambda = \frac{4ke^{-k(t-u)}}{\sigma^2 (1-e^{-k(t-u)})}r(u)
\end{equation}
Equivalently, one could write:
\begin{equation}
    \Pb\left(r(t)\leq y \mid r(u)\right) = F_{{\chi'}_\nu^2(\lambda)}\left(\frac{4 k y}{\sigma^2 (1-e^{-k(t-u)})}\right)
\end{equation}
Thus, due to these simple properties, it is possible to simulate the CIR model exactly on a discrete time grid, provided that it is possible to sample from a noncentral chi-square distribution \cite{scott}.

The CIR process can be sampled exactly on a discrete time grid $r(t_i)$, $t_i\in\Tc$, provided a sampling technique for noncentral chi-square distribution. This sampling method revolves around first generating a Poisson random variable $N$ and then, conditional on $N$, sampling a chi-square random variable with $\nu+2N$ degrees of freedom \cite{glasserman}. Fortunately, Scipy\cite{Scipy}, with the module `Stats`, has already implemented all the necessary distributions. The algorithm works as shown in Algorithm \ref{algo:cir_process}, where the random variable ${\chi'}^2_{d}(nc)$ is sampled using \texttt{scipy.stats.ncx2}.

\begin{algorithm}
    \caption{CIR process simulation}
    \begin{algorithmic}
        \For{$i:1,\dots,n$}
        \State $dt_i \gets t_i - t_{i-1}$
        \State $c \gets \sigma^2 \left(1-e^{-kdt}\right)/\left(4k\right)$
        \State $d \gets (4\theta k)/(\sigma^2)$
        \State $nc \gets r(t_{i-1})\cdot\left(e^{-kdt}\right)/c$
        \State $r(t_i) \gets c\cdot\left({\chi'}^2_{d}(nc)\right)$
        \EndFor
    \end{algorithmic}
\label{algo:cir_process}
\end{algorithm}

\begin{figure}[h]
    \centering
    \includegraphics[width=0.9\textwidth]{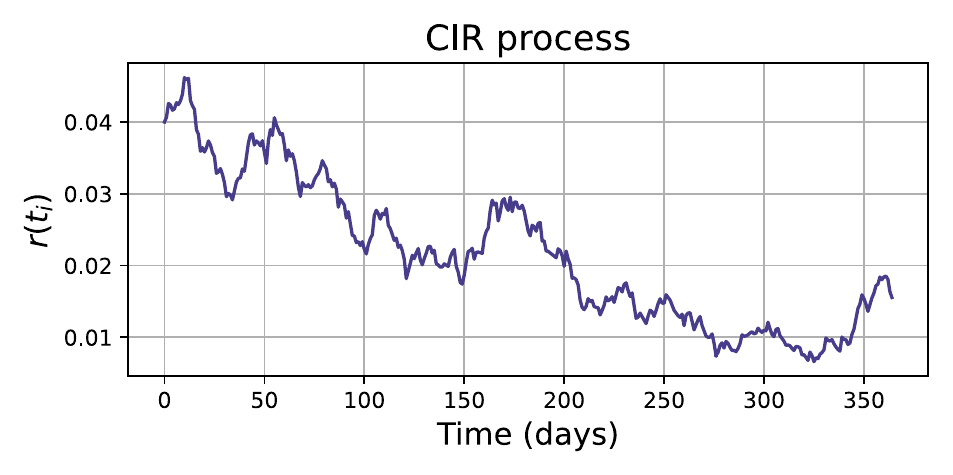}
    \caption{Simulation example of the CIR process with 1 year of length, obtained thorugh the Python implementation of algorithm \ref{algo:cir_process}.}
    \label{fig:cir_process_examples}
\end{figure}

Figure \ref{fig:cir_process_examples} shows a simulated realization of the CIR process.

The list of exact parameters used for the process is adopted from \cite{duffie} and is somewhat calibrated to exhibit certain market features. The actual parameters are:
\begin{equation*}
    k=0.6,\; \theta=0.04,\; \sigma=0.14
\end{equation*}

\subsection{Arrival time process}

Once the reference interest rate process is obtained, it is possible to simulate the arrival process of the contracts. Following \cite{giesecke_default_timings}, first, the continuous Cox process $\left\{\tilde{t}_k^{(ij)}\right\}$ is simulated. Over the fixed filtered probability space $\left(\Omega, \{\Fc_t\}_{t\geq0}, \Qb\right)$, consider a sequence of stopping times $\left\{\tilde{t}^{(ij)}_k\right\}$ that is strictly increasing and for which $\tilde{t}_0^{(ij)}=0$. Then denote by $\{N(t)\}_{t\geq0}$ the counting process given by:
\begin{equation}
    N(t)=\sum_{k\geq1}{\1_{\left\{\tilde{t}^{(ij)}_k\leq t\right\}}}
\end{equation}
Given that the stochastic intensity $\lambda_t$ follows an $\Fc_t$-adapted process, then Meyer \cite{meyer1971demonstration} proved that given the compensator $A$ that makes $M = N - A$ a \textit{local martingale}, $N$ is a standard Poisson process under a change of time defined through $A$. Thus, for a sequence $\{\xi_n\}_{n\in\Nb}$ of standard i.i.d. exponential random variables $\xi_n \sim \exp(\text{scale}=\chi)$, the arrival time of the counting process can be obtained through the time-change trick:
\begin{equation}
    \tilde{t}_k^{(ij)} = \inf_{t\geq0}{ \left\{ \int_0^t{\lambda_s ds \geq \xi_1 + \dots + \xi_k} \right\}}
\end{equation}
Therefore, the procedure works as follows:
\begin{algorithm}
    \caption{Cox process simulation}
    \begin{algorithmic}
    \State $\xi_1 \sim \exp(\text{scale}=\chi)$
    \State $\tilde{t}_1^{(ij)} \gets \inf_{t\geq0}{ \left\{ \int_0^t{\lambda_s ds \geq \xi_1} \right\}}$
        \While{$\sum_{l=1}^{n}{\xi_l}\leq T \leq \sum_{l=1}^{n+1}{\xi_l}$}
            \State $\xi_l \sim \exp(\text{scale}=\chi)$
            \State $\tilde{t}_l^{(ij)} \gets \inf_{t\geq0}{ \left\{ \int_0^t{\lambda_s ds \geq \sum_{j=0}^l \xi_j} \right\}}$
        \EndWhile
    \end{algorithmic}
\label{algo:Cox_process}
\end{algorithm}
Once the continuous process is obtained, the arrival times of the contracts is determined by transposing the Cox process onto the discrete grid through the intuitive rule:
\begin{equation}
    t_k^{(ij)} = \inf_{t\in\Tc}{\left\{ \tilde{t}_k^{(ij)} \leq t \right\}} 
\label{eq:Cox_discretize}
\end{equation}
\begin{figure}[h]
    \centering
    \includegraphics[width=0.9\textwidth]{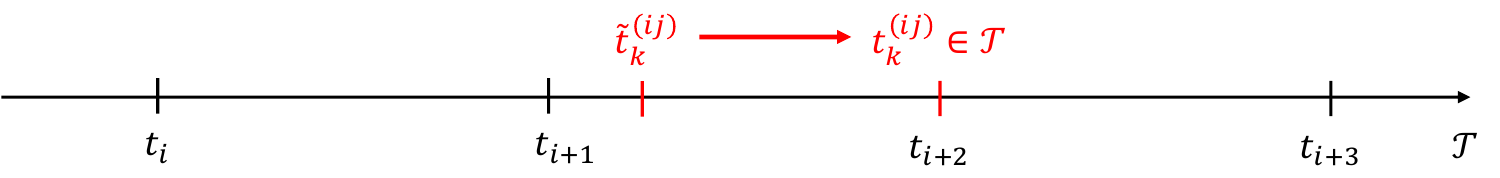}
    \caption{Sketch of the discretization of the continuous Cox process described in Eq. \eqref{eq:Cox_discretize}}
    \label{fig:Cox_discretized}
\end{figure}

\subsection{Graph simulation}
Once the previous parts are properly defined and implemented, simulating the graph is straightforward. The strategy adopted here is to firstly simulate the upper triangle of the edge tensor $E=\{E_t\}_{t\in\Tc}$, where the entries of $E$, $E_{ij}$ contain all the contracts between nodes $(ij)$ over the simulation horizon $[0,T]$ and are obtained as described in the previous sections. The lower triangle of $E$ is obtained simply by inverting $\delta_k^{(ij)}$ for each of the observed contracts. Subsequently, $E$ is batched in snapshots representing the graph at each time step containing all the relevant quantities, including $M(t)$, which is then used as label of the regression problem. The theoretical best predictor, or \textit{benchmark}, as well as all the other quantities that contain an expectation over the reference rate process, are obtained through a Monte Carlo estimation using a number $N=10^4$ of generated CIR paths.
Finally, the graph snapshots are stored in \texttt{Pythorch Geometric} \cite{Pytorch_geometric} data structures that are memory-efficient and implemented for the use of graph neural networks.
\begin{figure}[h]
    \centering
    \includegraphics[width=1\textwidth]{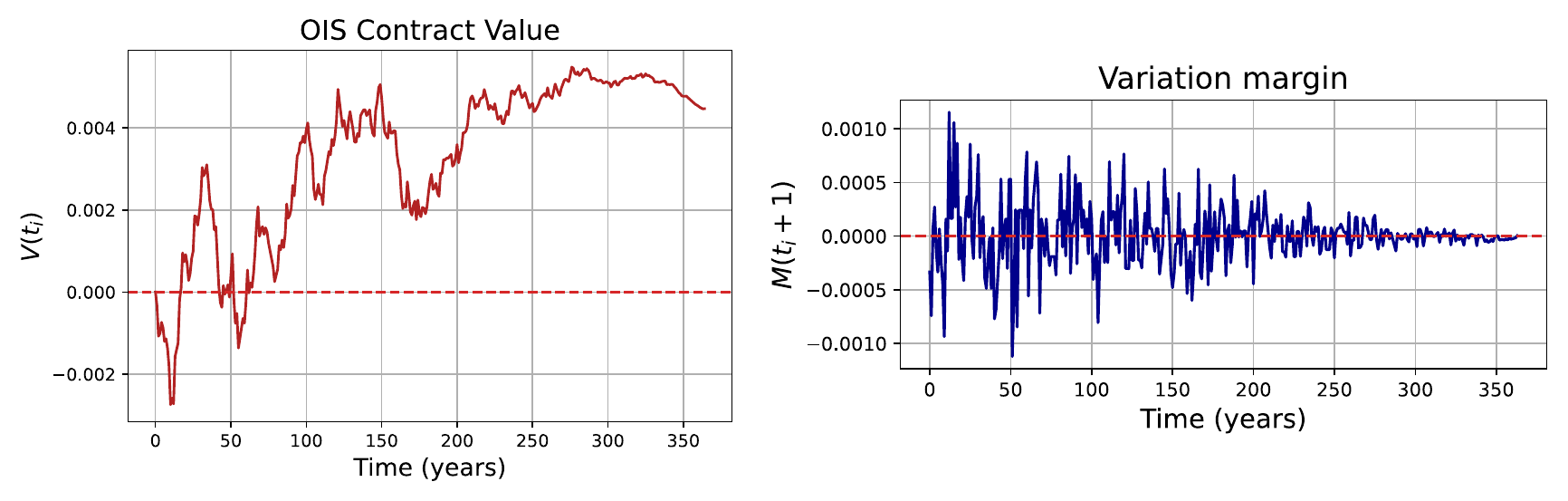}
    \caption{Illustrative figure representing the simulated $V(t_i)$ of a contract (plot on the left), and the associated margin $M(t_i)$ (right plot). }
\end{figure}

\section{Training}
\label{app:training}

Although the two modules require different datasets, they undergo the same windowing process.

This procedure, depicted in Fig. \ref{fig:data_windowing}, involves generating smaller sequences or \emph{windows}, from the original $T$-steps simulation. 

For the GNN module, the windows consist of snapshots of the discrete temporal graph stored in a suitable \texttt{PyTorch Geometric} \cite{Pytorch_geometric} format, containing both node features $X_t$ and the adjacency matrix $A_t$ in Pytorch's sparse COO format. Differently, windows for the Pricing module comprise the aformentioned contract matrices (see Fig. \ref{fig:contract_matrix}). The `.forward()` method of the model is implemented to skip empty columns that were padded with zeros, ensuring standardized-shaped matrices.

In the $N$-nodes, $m$-steps ahead setting, the complete model is trained by minimizing the MSE loss over the average of the loss computed at each of the intermediate $m$-steps, including the target one at time $t_{l+m}$. The weights of the two models are updated simultaneously, enabling the pricing module to enhance contract pricing given the process's intensity node embedding, while the GNN module learns to provide improved embeddings to the pricing module. It's important to note that while the model's performance should be compared to the best theoretical predictor in Eq. \ref{eq:benchmark}, the model must be trained using ground-truth labels $M(t_{l+m})$ obtained from the simulation. 

Overall, the model is trained using the Adam optimizer \cite{adam} and a learning scheduler that gradually decreases the learning rate as the model converges towards a local minimum of the loss landscape. The implemented code supports CUDA acceleration and was executed on a cluster equipped with various cutting-edge GPU hardware components. 

The complete list of technical specifications is provided below.

\subsection{Hardware}
In the following, the technical specifications of the used computational cluster is reported.
\begin{table}[H]
    \centering
    \caption{Clients Specifications}
    \label{tab:server_specs}
    \begin{tabular}{|l|l|l|}
        \toprule
        \textbf{Name} & \textbf{CPU / \textcolor{green}{GPU}} & \textbf{RAM} \\
        \hline
        ada-16 & 2x Intel Xeon E5-2697 v2 (12 Cores) 2.70GHz &  128 GB  \\
        ada-17 & 2x Intel Xeon E5-2697 v2 (12 Cores) 2.70GHz & 256 GB  \\
        ada-18 & 2x Intel Xeon E5-2697 v2 (12 Cores) 2.70GHz & 256 GB  \\
        ada-19 & 2x Intel Xeon E5-2697 v2 (12 Cores) 2.70GHz & 256 GB  \\
        ada-20 & 2x Intel Xeon E5-2699 v4 (22 Cores) 2.20GHz & 512 GB  \\
        ada-21 & 2x Intel Xeon E5-2699 v4 (22 Cores) 2.20GHz & 512 GB  \\
        ada-22 & 2x Intel Xeon E5-2697 v4 (18 Cores) 2.3 GHz & \\
               &  \textcolor{green}{1x NVIDIA GeForce GTX 1080 Ti (12GB)} & \\
               &  \textcolor{green}{1x NVIDIA GeForce GTX 1080 (8GB)} & 256 GB \\        
        ada-23 & 2x Intel Xeon Gold 6148 (20 Cores) 2.4 GHz & \\
               &  \textcolor{green}{4x NVIDIA GeForce GTX 1080 Ti (12 GB)} & 256 GB  \\
        ada-24 & 2x Intel Xeon Gold 6252 (24 Cores) 2.1 GHz & \\
               &  \textcolor{green}{2x NVIDIA GeForce RTX 2080 Ti (12 GB)} & 512 GB  \\
        ada-25 & 2x Intel Xeon Gold 6254 (18 Cores) 3.1 GHz & \\
               &  \textcolor{green}{2x NVIDIA GeForce RTX 2080 Ti (12 GB)} & 512 GB \\
        ada-26 & 1x AMD EPYC 7742 (64 Cores) 2.25GHz & \\
               &  \textcolor{green}{2x NVIDIA GeForce RTX 3090 (24 GB)} & 512 GB \\
        ada-27 & 1x AMD EPYC 7763 (64 Cores) 2.45GHz & \\
               &  \textcolor{green}{2x NVIDIA GeForce RTX 4090 (24 GB)} & 512 GB  \\
        ada-28 & 2x AMD EPYC 7313 (16 Cores) 3.0GHz & \\
               &  \textcolor{green}{4x NVIDIA GeForce RTX 4090 (24 GB)} & 1’024 GB \\
        \bottomrule
    \end{tabular}
\end{table}

\subsection{Model capacity}

\paragraph{Pricing module} The pricing module comprises an LSTM along with a FFNN. The following are the sizes of each layer:

\begin{table}[H]
    \centering
    \caption{Pricing Module specifications}
    \label{tab:pricing_module_specs}
    \begin{tabular}{|l|l|}
    \toprule
        \textbf{Layer name} & \textbf{input / output size}  \\
        \hline
        torch.nn.LSTM & (6/15) \\
        torch.nn.Linear & (15 $\times$ 2 +1 / 512) \\
        torch.nn.ReLU & / \\
        torch.nn.Linear & (512 / 512) \\
        torch.nn.ReLU & / \\
        torch.nn.Linear & (512, 6) \\
        torch.nn.ReLU & (6,1) \\
    \bottomrule
    \end{tabular}
\end{table}

\paragraph{GNN module} The GNN module comprises a GC-LSTM module. The following parameters were adopted throughout the study:
\begin{table}[H]
    \centering
    \caption{GNN Module specifications}
    \label{tab:GNN_module_specs}
    \begin{tabular}{|l|l|}
    \toprule
        \textbf{Spec name} & \textbf{value}  \\
        \hline
        in-channels & 1 \\
        
        out-channels & 15 \\
        
        K & 2 \\

    \bottomrule
    \end{tabular}
\end{table}

\subsection{Training specifications}

In addition to the model's parameters, there are other important parameters used during training. The actual parameters may differ according to the specific run, however here are reported the parameters used for generating Fig. \ref{fig:3_nodes_2_steps}.
\begin{table}[H]
    \centering
    \caption{Training specifications used for generating Figure \ref{fig:3_nodes_2_steps}}
    \label{tab:training_specs}
    \begin{tabular}{|l|l|}
    \toprule
        \textbf{Spec name} & \textbf{value}  \\
        \hline
        $\alpha$ & $0.6$ \\
        $b$ & $0.04$ \\
        $\sigma$ & $0.14$ \\
        $v_0$ & $0.04$ \\
        years & $60$ \\
        $\gamma$ & $3$ \\
        $\eta$ & $-4$ \\
        $\theta$ & $20$ \\
        $\beta$ & $5$ \\
        steps-ahead & $2$ \\
        lookback & $5$ \\
        train-split & $0.8$ \\
        initial-lr & $5e-3$ \\
        n-epochs & $20\cdot10^3$ \\
        patience & $50$ \\
        scheduler-frequency & $130$ \\
        scheduler updates & $60$ \\
        scheduler $\gamma$ & $0.9$ \\
        validation-every & $20$ \\
        batch size & $500$ \\
    \bottomrule
    \end{tabular}
\end{table}










\end{document}